\documentclass{tlp}

\usepackage{url}
\usepackage{amsmath}
\let\proof\relax

\usepackage{amssymb,amsthm,mathrsfs,amsfonts,dsfont} 
\usepackage{xcolor,colortbl}
\usepackage{mdframed}
\usepackage{graphicx}
\usepackage{graphics} 
\usepackage{epsfig} 
\usepackage{multirow}
\usepackage{xspace}

\usepackage{listings}
\usepackage{asplisting}
\usepackage[utf8]{inputenc}
\usepackage{multirow}
\usepackage{comment}
\usepackage{todonotes}
\usepackage{subcaption}

\definecolor{asparagus}{rgb}{0.53, 0.66, 0.42}

\lstnewenvironment{asp}[1][]
{
	\lstset{
		language=asp,
		showstringspaces=false,
		formfeed=\newpage,
		tabsize=4,
		commentstyle=\color{asparagus},
		basicstyle=\ttfamily\scriptsize,
		keywordstyle=\color{blue},
		numbers=left,
		breaklines=true,
		literate={~} {$\sim$}{1},
		#1
	}
}
{
}

\newcommand{\code}[1]{\texttt{#1}}

\theoremstyle{definition}

\theoremstyle{remark}

\newcommand{\clingo}{\textsc{clingo}\xspace}

\newtheorem{definition}{Definition}

\begin{document}

\title{Solving Rehabilitation Scheduling problems via a Two-Phase ASP approach\footnote{This paper is an extended and revised version of a conference paper appearing in the proceedings of the RuleML+RR 2021 conference \citep{DBLP:conf/ruleml/CardelliniNDGGM21}.}}

\righttitle{Solving Rehabilitation Scheduling problems via a Two-Phase ASP approach}

\jnlPage{1}{8}
\jnlDoiYr{2021}
\doival{10.1017/xxxxx}

\begin{authgrp}
    \author{\gn{Matteo} \sn{Cardellini}}
    \affiliation{Polytecnic of Torino, Torino, Italy\\University of Genova, Genova, Italy}
    \author{\gn{Paolo} \sn{De Nardi}}
    \affiliation{ICS Maugeri, Italy}
    \author{\gn{Carmine} \sn{Dodaro}}
    \affiliation{DeMaCS, University of Calabria, Rende, Italy}
    \author{\gn{Giuseppe} \sn{Galat\`a}}
    \affiliation{SurgiQ srl, Italy}
    \author{\gn{Anna} \sn{Giardini}}
    \affiliation{ICS Maugeri, Italy}
    \author{\gn{Marco} \sn{Maratea}}
    \affiliation{DIBRIS, University of Genova, Genova, Italy\\DeMaCS, University of Calabria, Rende, Italy}
    \author{\gn{Ivan} \sn{Porro}}
    \affiliation{SurgiQ srl, Italy}
\end{authgrp}
\lefttitle{Cardellini et al.}

\history{\sub{xx xx xxxx;} \rev{xx xx xxxx;} \acc{xx xx xxxx}}

\maketitle

\begin{abstract}

A core part of the rehabilitation scheduling process consists of planning rehabilitation physiotherapy sessions for patients, by assigning proper operators to them in a certain time slot of a given day, taking into account several legal, medical and ethical requirements and optimizations, e.g.,  patient's preferences and operator's work balancing. Being able to efficiently solve such problem is of upmost importance, in particular after the COVID-19 pandemic that significantly increased rehabilitation's needs.

In this paper, we present a two-phase solution to rehabilitation scheduling based on Answer Set Programming, which proved to be an effective tool for solving practical scheduling problems. We first present a general encoding, and then add domain specific optimizations. Results of experiments performed on both synthetic and real benchmarks, the latter provided by ICS Maugeri, show the effectiveness of our solution as well as the impact of our domain specific optimizations.
Under consideration in Theory and Practice of Logic Programming (TPLP).
\end{abstract}


\section{Introduction}

The rehabilitation scheduling process consists mainly of planning daily patients’ physiotherapy sessions inside a rehabilitation institute, that hereafter we refer to as Rehabilitation Scheduling Problem (RSP) \citep{Huang2012DecisionSS,huynh_hybrid_2018,Li_2021,schimmelpfeng_decision_2012}. Hospitals that may profitably make a practical use of such scheduling, including those managed by ICS Maugeri\footnote{\url{https://www.icsmaugeri.it/}.}, which will provide benchmarks in this paper, deal with up to hundreds of patients with a team of just few tens of physiotherapists; so, it is of paramount importance to be able to assign patients to operators, i.e., physiotherapists, efficiently. 
A recent article by \citeauthor{cieza_global_2020} (\citeyear{cieza_global_2020}) found that 2.41 billion people could benefit from rehabilitation services.
This finding means that almost one third of the current population in the world needs rehabilitation at some point during the course of their lives due to disease or injury; further, this number is predicted to trend upward given the current
demographic and health shifts. In addition, there is emerging evidence that many of the people affected by the COVID-19 pandemic have long-term consequences regardless of the disease severity or length of hospitalisation, thus further increasing the demand for rehabilitation services globally.

The RSP is subject to several constraints, i.e., legal, medical and ethical, that need to be taken into consideration in order to find a viable schedule. For example, the main constraints that have to be dealt with are the maximum capacity of rehabilitation gyms, the legal working time and rest periods for operators, and the minimum durations of physiotherapy sessions. Moreover, several preferences shall be considered, e.g., due to clinical and organizational reasons it is often best for a patient to be treated as often as possible by the same operator and at the same time slot; also, rehabilitation professionals' work balancing needs to be taken into proper account.

In this paper, we 
present a solution to the RSP based on Answer Set Programming (ASP)~\citep{DBLP:journals/ngc/GelfondL91,DBLP:journals/amai/Niemela99,baral2003,DBLP:journals/cacm/BrewkaET11}, which proved to be an effective tool for solving practical scheduling problems \citep{DBLP:journals/tplp/GebserOSR18,DBLP:journals/tplp/RiccaGAMLIL12,DBLP:conf/lpnmr/DodaroM17,DBLP:journals/tplp/DodaroGGMMP21}, thanks also to the availability of efficient ASP solvers.
The solution is designed as a two-phase encoding (Section \ref{sec:enc}):
the first phase, called {\em board}, deals with the problem of assigning a physiotherapist to every patient considering the total working time of the physiotherapist and the minimum mandatory time of rehabilitation sessions. In the second phase, called {\em agenda}, a start and end time of every rehabilitation session is defined given the assignment among patients and physiotherapists found in the first phase. Our two-phase solution is not guaranteed to find the best possible overall solution, but has been designed in this way because: $(i)$ it simplifies the overall encoding and its practical use, and $(ii)$ it mimics how schedules have been computed so far (in a non-automatic way) by ICS Maugeri and gives freedom to physiotherapists' coordinators to perform any desired manual change. In fact, coordinators have specifically requested to have the possibility to manually change some patient-operator assignments (the output of the \textit{board}) before sending it to the \textit{agenda} phase. Even if this manual change is seldom made, it gives the coordinators a sense of control and the possibility to introduce human knowledge and expertise in the scheduling. It is important to acknowledge that this tool is not advertised and sold as a medical device, but it is a tool for supporting the decision of the coordinators. For this reason, it is legally mandatory for the coordinators to have a more granular control (and responsibility) over the decisions. We first tested (Section \ref{sec:res}) our encoding on real scenarios from ICS Maugeri related to the daily scheduling of neurological patients in two of their rehabilitation institutes in the North of Italy, namely Genova Nervi and Castel Goffredo. 
In the analysis, we decided to limit the run-time of the ASP solver \clingo \citep{DBLP:journals/ai/GebserKS12} to only $30s$ (while in production the cut-off is set to $5$ minutes): This narrow time limit allows for running much more experiments and having a more significant comparison with the different optimization algorithms in \clingo. 
Then, given that ICS Maugeri is planning to instrument with automated techniques other, possibly larger, institutes in addition to Genova Nervi and Castel Goffredo, we generated a wide set of synthetic benchmarks, whose parameters are inspired by real data. We made a wide experimental evaluation, and statistically confronted synthetic and real data results using classification decision tree methods \citep{quinlan1986induction}, with the aim of predicting the behaviour of our solution on such larger institutes. Results show that the accuracy is high, so our synthetic benchmarks appear significant to indicate a possible behavior on real data coming from other institutes with other parameters and similar characteristics. As a side effect, this analysis also outlines the features of the problems that affect the results mostly. Finally, with the aim of further improving the results, and lower the still remaining percentage of instances that could not be solved, we added domain specific optimizations to our encoding (Section \ref{sec:opt}): results of the improved encoding show that we are now able to find a solution, even if not always optimal, to every instance within the time limit. The paper is completed by an informal description of the RSP in Section \ref{sec:descr} and by discussing related work and presenting conclusions in Section \ref{sec:rel} and \ref{sec:conc}, respectively.

\section{Problem Description}
\label{sec:descr}

In this section we describe the problem we face in four paragraphs. First, we present the general description of the problem, then the data that characterize the main elements of the problem, followed by the requirements of the phases. The last paragraph shows a solution schedule.

\paragraph{\bf General description.}
The delivery of rehabilitation services is a complex task that involves many healthcare professions such as physicians, physiotherapists, speech therapists, psychologists and so on. In particular, physiotherapists are the ones who spend most of their time with patients and their sessions constitute the core of the daily agenda of the patient, around which all other commitments revolve. For this reason, this article is focused on scheduling the physiotherapy sessions in the most efficient way, optimising the overall time spent with the patient.

The agenda for the physiotherapy sessions is computed by the coordinator of the physiotherapists. This process is repeated on a daily basis in order to take into account any change in the number and type of patients to be treated, and the number of operators available. Up until recently, this computation has been performed manually by coordinators, without any decision support. 

The usual scheduling practice entails two subsequent phases resulting in the computation of a board and an agenda, that we herewith describe. In short, the first phase, called {\em board}, deals with the problem of assigning a physiotherapist to every patient, keeping track of the total working time of the operator and the minimum mandatory time of rehabilitation sessions. In the second phase, called {\em agenda}, a start and end time of every rehabilitation session is computed, given the assignment among patients and operators found in the first phase.

In more details, in the board phase, we ensure that the working hours of operators are respected by counting their total working time, in minutes, and assigning patients to each operator in such a way that the cumulative time of all their sessions remains below the operator's total working time. 
In this phase, patient-operator assignment preferences, expressed by the coordinator before the start of the scheduling procedure, are taken into account and respected as far as possible. In the agenda phase, given an assignment found by the board, every patient-operator session is assigned a starting and ending time, respecting the more granular working hours of the operators and the times in which the patients are unavailable. At this stage, the location in which the rehabilitation session is performed is also considered. A location, either a gym or the room of the patient, is assigned to the session, according to the clinical needs of the patient. The choice of the gym is carried out by considering the maximum number of simultaneous sessions allowed inside the gym and has to be made among a subset of gyms which are located at the same floor as the room of the patient, in order to avoid elevators and stairs that can result in discomfort to patients and slowness which can quickly congest the hospital. In this phase, time preferences for each patient are also considered: in fact, plans in which the sessions are performed closer to the desired time of the patients are to be preferred to others.

\paragraph{\bf Instance description.} In this paragraph we describe the main elements of our problem in more details, namely patients, operators and sessions, as well as the constraints and preferences entailed by the board and agenda phases.

\paragraph{Patients.}
\label{subsec:patient}
Patients are characterized by their:
\begin{itemize}
    \item type (Neurological, Orthopaedic, COVID-19 Positive, COVID-19 Negative, Outpatient\footnote{A person who goes to a hospital for a daily treatment, without staying the night.}), 
    \item aid needs, i.e., if they need specific care or not (e.g., if they need to be lifted),
    \item payment status (full payer or in charge of the National Healthcare Service),
    \item forbidden times, i.e., the time intervals when the patient cannot be scheduled,
    \item ideal time, i.e., the preferred scheduled timeslot in which the session should take place, expressed by the coordinator,
    \item preferred operators, i.e., the list of physiotherapists, ordered by priority, the patient can be assigned to,
    \item overall minimum length, i.e., the minimum amount of care time that the patient is guaranteed to be scheduled,
    \item sessions, i.e., the list of sessions to be scheduled.
\end{itemize}

\paragraph{Operators.}
\label{subsec:operator}
Physiotherapists, which will be called operators from now on, are characterized by their:
\begin{itemize}
    \item qualifications, i.e., patient's types the operator can treat, 
    \item operating times, i.e., the part of the operator's working times dedicated to the direct care of the patients. The operating times are usually split in morning and afternoon shifts.
\end{itemize}
Moreover, each operator has a limit on the number of patients of a specific type to treat.

\paragraph{Sessions.}
\label{subsec:session}
The coordinator, in accordance with the rehabilitation program set by the physician, determines the daily activities of the patient. These activities can be performed in one or two therapy sessions, in the latter case one session will be scheduled in the morning and the other one in the afternoon shift.

\begin{figure}[t]
\includegraphics[width=\textwidth]{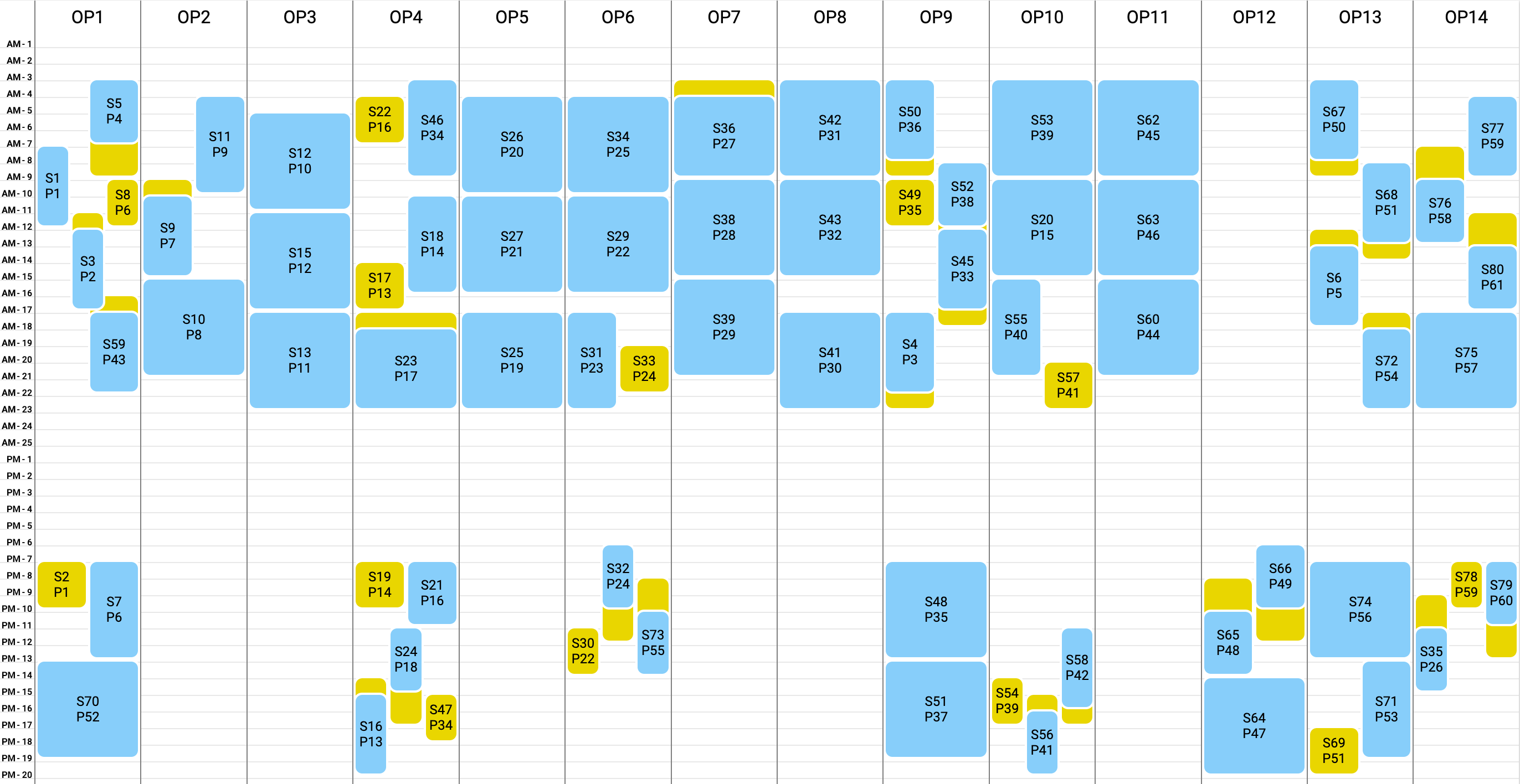}
\caption{Result of the scheduling of the agenda in a real case scenario in the hospital of Genova Nervi. Light blue (yellow) squares represent time units in which the sessions will be performed in an individual (supervised) fashion. The ticks on the left keep track of the period (morning or afternoon) and time slot in which the session will start or end.}
\label{fig:example-agenda-scheduling}
\end{figure}

Each session can be delivered to patients either individualized (“one-on-one” sessions) or supervised  (one therapist supervising more patients at the same time, each patient carrying out their personal activity independently). 
It must be noted that while operators deliver one-on-one therapy to one patient, they can supervise other patients. When the operators are particularly overbooked, their one-on-one sessions can be partially converted to supervised ones. These mixed sessions can either start with a supervised part and then continue with the one-on-one part, or vice-versa, or even start and end with a supervised part with a middle one-on-one session. Obviously, an operator can supervise different patients only if their sessions are located at the same place. In the next paragraphs, when defining the \textit{agenda}, Figure \ref{fig:example-agenda-scheduling} will graphically explain the semantic of a mixed session.

\noindent The characteristics of the sessions are:
\begin{itemize}
    \item delivery mode (one-on-one, supervised),
    \item minimum one-on-one length, i.e., the minimum length of the session guaranteed to be delivered one-on-one,
    \item ideal overall length, i.e., the overall length of the session including the one-on-one and supervised parts,
    \item optional status, i.e., if the session can be left out of the schedule in case of overbooked operators,
    \item forced time, i.e., the time when the session must be scheduled; if empty, the session is placed as close as possible to the patient's preferred time,
    \item location, i.e., the place where the session must be delivered. 
\end{itemize}

\paragraph{\bf Constraints of the phases.} The requirements that the two phases entail are reported in the following sub-paragraphs.

\paragraph{Board.}
\label{subsec:board}
In the board phase, all patients are assigned to an available operator, according to the following criteria:
\begin{itemize}
    \item compatibility between patient and operator, depending on the patient's type and operator qualifications, the patient's forced time, if any, and the operator working times, by also checking if the operator has enough time to provide the guaranteed overall minimum length and minimum one-on-one length to each patient and session,
    \item the patients should be fairly distributed among all available operators, taking into account their type, aid needs and payment status,
    \item the patients should be assigned to the operators respecting as much as possible their preferred operators list, which considers primarily the choices of the coordinator and secondarily the history of the past assignments.
\end{itemize}


\paragraph{Agenda.}
\label{subsec:agenda}

The results of the board phase can be revised (e.g., in special cases, the coordinator can override the preferred operators list and force an assignment of a patient to an operator regardless of all other considerations) and, if necessary, manually modified by the coordinator. Once the coordinator is satisfied with the board, it is possible to proceed to the agenda scheduling, using the approved board as input. 
The criteria for the agenda phase are:
\begin{itemize}
    \item compliance with the forced time of the session, if specified,
    \item two sessions of the same patient must be assigned in different shifts,
    \item compliance with the minimum one-on-one length of the session,
    \item no overlap between two one-on-one sessions (or the one-on-one part of the session if mixed) assigned to the same operator, 
    \item observance of the maximum capacity of the locations (1 for each room, varying for the gyms),
    \item respect of the minimum cumulative time that the patient should be treated among all the sessions,
    \item respect of the one-on-one minimum session length,
    \item compliance with the forbidden times of the patient,
    \item sessions can only be scheduled within the working times of the operator,
    \item the start time of each session should be as close as possible to the preferred time, either specified by the coordinator or inferred from previous schedules,
    \item for mixed sessions, the one-on-one part should be maximized,
    \item the largest possible number of optional sessions should be included,
    \item the overall length, including the one-on-one and supervised parts in case of mixed sessions, should be as close as possible to the ideal overall length specified by the coordinator.
\end{itemize}

\paragraph{\bf Scheduling example.} As previously stated, the output of the \textit{agenda} phase is a start time and duration of every session during the day. Since, in ICS Maugeri, the scheduling was already performed with a  timeslot of $10$ minutes, we decided to keep this discretization in place. For this reason, the sessions can start every 10 minutes in the working hours of the hospital (8AM-12AM in the morning, 1:30PM - 4PM in the afternoon) and last a multiple of 10 minutes. Figure \ref{fig:example-agenda-scheduling} shows the scheduling of the agenda in a real case scenario in the hospital of Genova Nervi. Light blue squares represent time units in which the sessions will be performed in an individual fashion, and yellow squares represent time units of sessions in which the patient will be dealt in a supervised mode. Ticks on the left side of the figure describe the period (AM = Morning, PM = Afternoon) and the number we associate to each timeslot. As it can be seen in the first column, Operator 1 (\texttt{OP1}) deals with the first session (\texttt{S5}) as a mixed session: the session starts in individual one-to-one mode with all the attention of the operator focused on the patient (\texttt{P4}). After 4 time slots (i.e., 40 minutes, the session minimum time) the operator moves the attention to another patient (\texttt{P1} performing session \texttt{S1}) while the previous patient finishes the session, in the same room, on its own. In a more practical way, in the first 40 minutes the operator helps the patient in performing exercises which could not be performed alone in a correct way. In the remaining 20 minutes, which are still very beneficial to the patient, the patient performs the exercises that can be done alone while the operator is working with another patient. Being in the same room, the operator can still intervene if the supervised patient needs correction. Some sessions (e.g., session \texttt{S17} of patient \texttt{P13} performed in the morning by operator \texttt{O4}) are performed in a complete supervised fashion since they are additional secondary sessions which are medically beneficial to the patient but not mandatory (e.g., patient \texttt{P13} already performs session \texttt{S16} in an individual fashion with operator \texttt{O4} in the afternoon).

\section{A Two-Phase ASP Encoding for the RSP}
\label{sec:enc}

In the following, we assume the reader is familiar with syntax and semantics of ASP.
Starting from the specifications in the previous section,
here we present the ASP encoding, based on the input language of {\sc clingo} \citep{DBLP:conf/iclp/GebserKKOSW16}. For details about syntax and semantics of ASP programs, we refer the reader to \citep{DBLP:journals/tplp/CalimeriFGIKKLM20}.

\subsection{Board encoding}
\paragraph{Data Model.} The input data is specified by means of the following atoms:
\begin{itemize}
\item Instances of \texttt{patient(P)}, \texttt{operators(O)}, and \texttt{type(T)} represent the identifiers of patients, operators, and the different types of patients that can be visited, respectively, where \texttt{P} and \texttt{O} are numbers, whereas \texttt{T} is of the form \texttt{value-needs-status}, where \texttt{value} can be 
\textit{neurologic}, \textit{orthopaedic}, \textit{covid-19-positive}, \textit{covid-19-negative}, or \textit{outpatient}; \texttt{needs} can be \textit{lifter} or \textit{nolifter}, and \texttt{status} can be \textit{payer} or \textit{free}.
For instance, \texttt{neurologic-lifter-payer} indicates that the patient needs a neurological treatment, must be lifted, and the treatment must be paid.
Moreover, a fictitious operator with \texttt{ID} equals to -1 is included in the list of all the operators, and it is needed to intercept all patients that cannot be assigned to other operators (like a \textit{catch-all}\footnote{This is because, for practical reasons, we always want to have a solution.}).

\item Instances of \texttt{operator\_contract(ID,TIME,MAX)} represent the contract of the operator with the identifier \texttt{ID}, and include the quantity of time (in time units) the operator works in a day (\texttt{TIME}), 
and the maximum number of patients the operator can visit during the day (\texttt{MAX}).

\item Instances of \texttt{operator\_limit(ID,T,VALUE)} represent the maximum number of patients (\texttt{VALUE}) of type \texttt{T} the operator with identifier \texttt{ID} can visit.
The operator with \texttt{ID} equals to -1 has no patients limit.

\item Instances of \texttt{patient\_data(ID,T,DUR)} represent the data associated to the patient with the identifier \texttt{ID}, and include the type of the patient (\texttt{T}), and the minimum cumulative time of all sessions of the patient during the day (\texttt{DUR}).

\item Instances of \texttt{patient\_session(ID,MIN,LOC)} represent a rehabilitation session that the patient with identifier \texttt{ID} needs to perform during the day. The session is characterized by a minimum length for the session in time units (\texttt{MIN}), and the location of the session (\texttt{LOC}).

\item Instances of \texttt{patient\_preference(ID,OP,W)} represent the preference of the patient with identifier \texttt{ID} to be treated by the operator with identifier \texttt{OP}, where \texttt{W} specifies the weight of the preference. 

\item Similarly, instances of \texttt{history\_preference(ID,OP,W)} represent the preference of the patient based on the history of previous sessions in previous days.
\end{itemize}
The output is an assignment represented by atoms of the form \texttt{assignment(OP, PAT)}, stating that patient \texttt{PAT} will be treated by operator \texttt{OP}.

\begin{figure}[t!]
    \centering
    \begin{asp}
$\label{enc1:guess}${assignment(OP, PAT) : operator(OP)} = 1 :- patient(PAT).
$\label{enc1:unique}$uniqueLocationLength(OP,PAT,DUR) :- assignment(OP,PAT), patient_session(PAT,_,LOC), patient_data(PAT,_,DUR), #count{ID:patient_session(ID,_,LOC), assignment(OP,ID)} < 2.
$\label{enc1:same}$sameLocationLength(OP,PAT,DUR) :- assignment(OP,PAT), patient_session(PAT,DUR,LOC), #count{ID:patient_session(ID,_,LOC), assignment(OP,ID)} > 1.
$\label{enc1:constrtime}$:- operator_contract(OP,TIME,_), #sum{U,PAT:uniqueLocationLength(OP,PAT,U); S, PAT:sameLocationLength(OP,PAT,S)} > TIME.
$\label{enc1:constrpatients}$:- operator_contract(OP,_,N), #count{PAT:assignment(OP,PAT)} > N.
$\label{enc1:constrpatientstype}$:- operator_limit(OP,T,N), #count{PAT:assignment(OP,PAT), patient_data(PAT,T,_} > N.
$\label{enc1:weakprefpatient}$:~ #sum{W, PAT:assignment(OP,PAT), patient_preference(PAT,OP,W)} = N. [N@3]
$\label{enc1:weakfakeoperator}$:~ #count{PAT: assignment(-1, PAT)} = N. [N@2]
$\label{enc1:weakprefhistory}$:~ #sum{W, PAT:assignment(OP,PAT), history_preference(PAT,OP,W)} = N.  [N@1]
    \end{asp}
    \caption{ASP Encoding for the board problem.}
    \label{encoding:board}
\end{figure}

\paragraph{Encoding.} The related encoding is shown in Figure \ref{encoding:board}, and is described in the following. To simplify the description, the rule appearing at line $i$ in Figure \ref{encoding:board} is denoted with $r_i$. Rule $r_{\ref{enc1:guess}}$ ensures that each patient is assigned to exactly one operator.
Rules $r_{\ref{enc1:unique}}$ and $r_{\ref{enc1:same}}$ are used to define if the session between a patient and an operator will be performed individually in a single location ($r_{\ref{enc1:unique}}$), or it will be executed in the same location of another session ($r_{\ref{enc1:same}}$), by creating two auxiliary atoms \texttt{uniqueLocationLength(OP,PAT,DUR)} and \texttt{sameLocationLength(OP,PAT,DUR)} that represent the duration \texttt{DUR} in time slots of the session between operator \texttt{OP} and patient \texttt{PAT} performed in a single or same location, respectively, used in the next rule.
Rule $r_{\ref{enc1:constrtime}}$ ensures that the time required by the patients assigned to an operator does not exceed the maximum time of her/his contract.
Rule $r_{\ref{enc1:constrpatients}}$ ensures that each operator does not exceed the maximum number of patients to visit during the day.
Rule $r_{\ref{enc1:constrpatientstype}}$ is similar to the previous one, but in this case the limits are imposed according to the type of the patient.

Weak constraints from $r_{\ref{enc1:weakprefpatient}}$ to $r_{\ref{enc1:weakprefhistory}}$ are then used to provide preferences among different assignments.
In particular, $r_{\ref{enc1:weakprefpatient}}$ is used to maximize the assignments that fulfil the preferences of each patient.
Then, $r_{\ref{enc1:weakfakeoperator}}$ is used to minimize the number of patients that are assigned to the fictitious operator.
Finally, $r_{\ref{enc1:weakprefhistory}}$ is used to maximize the solutions that preserve assignments dictated by the history of previous sessions.

\subsection{Agenda encoding}
\label{sec:enc-agenda}
\paragraph{Data Model.} The following atoms constitute the input data:
\begin{itemize}
    \item Instances of \texttt{patient(ID,MIN)} represent a patient identified by \texttt{ID}, and a minimum rehabilitation session of \texttt{MIN} length in time units that the patient has to undertake during the day.
    \item Instances of \texttt{period(PER,OP,STA,END)} define the start (\texttt{STA}) and end  (\texttt{END}) time unit in the period \texttt{PER} (which can be \textit{morning} or \textit{afternoon}), which corresponds to the shift of the operator with identifier \texttt{OP}.
    \item Instances of \texttt{time(PER,OP,T)} define the time slots \texttt{T} during the period \texttt{PER} where the operator \texttt{OP} works. In particular, \texttt{T} ranges from \texttt{STA} to \texttt{END} defined by the above atom, i.e. \texttt{time} is defined as \lstinline[language=asp]|time(PER,OP,STA..END) :- period(PER,OP,STA,END)|.
    \item Instances of \texttt{location(ID,CAP,PER,STA,END)} represent a location (i.e., a gym or a room), with an identifier \texttt{ID}, a maximum capacity of \texttt{CAP}, which, during the period \texttt{PER}, is open from the time unit \texttt{STA} until \texttt{END}.
    \item Instances of \texttt{macro\_location(MLOC,LOC)} define that the location \texttt{LOC} is inside the macro-location \texttt{MLOC} (i.e., a floor).
    \item Instances of \texttt{session(ID,PAT,OP)} represent a session between the patient \texttt{PAT} and the operator \texttt{OP}, coming from the  \texttt{assignment(OP,PAT)} output of the board phase, to which a unique \texttt{ID} is added
    (to discriminate between \textit{morning} and \textit{afternoon} shifts).
    \item Instances of \texttt{session\_type(ID,OP,TYPE)} represent that the session with identifier \texttt{ID} assigned to operator \texttt{OP} is of type \texttt{TYPE} (which can be \textit{individual} or \textit{supervised}).
    \item Instances of \texttt{session\_macro\_location(ID,MLOC)} represent that the session with identifier \texttt{ID} has to be held in the macro-location \texttt{MLOC}.
    \item Instances of \texttt{session\_length(ID,MIN,IDEAL)} represent that the session \texttt{ID} has a minimum length (\texttt{MIN}) that has to be performed in individual, and an ideal length (\texttt{IDEAL}) that would be beneficial to the patient, but it is not mandatory to perform.
    \item Instances of \texttt{mandatory\_session(ID)} and \texttt{optional\_session(ID)} identify sessions that are mandatory and optional, respectively.
    \item Instances of \texttt{forbidden(PAT,PER,STA,END)} represent an unavailability of the patient \texttt{PAT} in the period \texttt{PER} from the time unit from \texttt{STA} to \texttt{END}.
    \item Instances of \texttt{session\_preference(ID,PER,START,TYPE)} represent the preference of the patient, stating that the session should be held during the period \texttt{PER} and it must start at the time unit \texttt{START}, where \texttt{TYPE} indicates if the preference is \textit{high} or \textit{low}.
\end{itemize}

The output is represented by atoms \texttt{start(ID,PER,T)}, \texttt{length(ID,PER,L)}, and \texttt{session\_location(ID,LOC)}, which indicate the start, length and location of each session, respectively. 
\begin{figure*}[t!]
    \centering
    \begin{asp}[name=timetable]
$\label{enc2:guessStartMandatory}$ {start(ID,PER,TS) : time(PER,OP,TS)} = 1 :- session(ID,_,OP), mandatory_session(ID).
$\label{enc2:guessStartOptional}$ {start(ID,PER,TS) : time(PER,OP,TS)} <= 1 :- session(ID,_,OP), optional_session(ID).
$\label{enc2:guessLength}$ {length(ID,PER,NL) : time(PER,OP,L), NL=L-ST, TS+NL <= END, NL>= MIN, NL<= IDEAL} = 1 :- start(ID,PER,TS), period(PER,OP,ST,END), session(ID,_,OP), session_length(ID,MIN,IDEAL).
$\label{enc2:guessSessionLocation}$ {session_location(ID,LOC): macro_location(MAC,LOC)} = 1 :- session_macro_location(ID,MAC).
$\label{enc2:guessBefore}$ {before(ID,NL): time(PER,OP,L), NL=L-ST, NL<=TS-ST} = 1 :- start(ID,PER,TS), period(PER,OP,ST,_), session(ID,_,OP).
$\label{enc2:guessAfter}$ {after(ID,NL): time(PER,OP,L), NL=L-ST, NL<=END-TS-LEN} = 1 :- start(ID,PER,TS), period(PER,OP,ST,END), length(ID,PER,LEN), session(ID,_,OP).
$\label{enc2:extstart}$ extstart(ID,PER,TS-LB) :- start(ID,PER,TS), before(ID,LB).
$\label{enc2:extlength}$ extlength(ID,PER,L+LA+LB) :- length(ID,PER,L), after(ID,LA), before(ID,LB).
$\label{enc2:individualSessionLocation}$ individual_session_location(ID,LOC,OP,MIN,IDEAL) :- session_type(ID,OP,individual), session_location(ID,LOC), session_length(ID,MIN,IDEAL). 
$\label{enc2:sessionAtTime}$ session_time(ID,OP,PL,PER,TS..TS+L-1) :- session(ID,_,OP), session_location(ID,PL), extstart(ID,PER,TS), extlength(ID,PER,L).
$\label{enc2:individualAssignmentsOverlap}$ :- start(ID,PER,TS), length(ID,PER,L), session_type(ID,OP,individual), start(ID2,PER,TS2), session_type(ID2,OP,individual), ID!=ID2, TS2>=TS, TS2<TS+L.
$\label{enc2:sessionPerPeriod}$ :- session(ID1,PAT,_), session(ID2,PAT,_), start(ID1,PER,_), start(ID2,PER,_), ID1!=ID2.
$\label{enc2:equalReduction1}$ :- individual_session_location(ID1,LOC,OP,MIN1,OPT1), length(ID1,PER,L1), individual_session_location(ID2,LOC,OP,MIN2,OPT2), length(ID2,PER,L2), OPT1-L1 <= OPT2-MIN2, OPT2-L2 <= OPT1-MIN1 , |OPT1 -L1 - OPT2 + L2| > 1.
$\label{enc2:equalReduction2}$ :- individual_session_location(ID1,LOC,OP,MIN1,OPT1), length(ID1,PER,L1), individual_session_location(ID2,LOC,OP,MIN2,OPT2), length(ID2,PER,L2), OPT1-L1 > OPT2-MIN2, L2 > MIN2.
$\label{enc2:equalReduction3}$ :- individual_session_location(ID1,LOC,OP,MIN1,OPT1), length(ID1,PER,L1), individual_session_location(ID2,LOC,OP,MIN2,OPT2), length(ID2,PER,L2), OPT1-L1 <= OPT2-MIN2, OPT2-L2 <= OPT1-MIN1, OPT2 < OPT1, OPT1-L1 < OPT2-L2.
$\label{enc2:ubiquity}$ :- session_time(ID,OP,PL,PER,T), session_time(ID2,OP,PL2,PER,T), ID != ID2, PL != PL2.
$\label{enc2:minTimeReserved}$ :- patient(PAT,MIN), #sum{LEN, ID: session(ID,PAT,_), extlength(ID,_,LEN)} < MIN.
$\label{enc2:gymCapacity}$ :- location(LOC,LIM,PER,ST,END), LIM>0, time(PER,_,T), T>=ST, T<END, #count{ID: session_time(ID,_,LOC,PER,T)} > LIM.
$\label{enc2:forbidden1}$ :- forbidden(PAT,PER,ST,_), session(ID,PAT,_), extstart(ID,PER,TS), extlength(ID,PER,L), ST>=TS, ST<TS+L.
$\label{enc2:forbidden2}$ :- forbidden(PAT,PER,_,END), session(ID,PAT,_), extstart(ID,PER,TS), extlength(ID,PER,L), END>TS, END<=TS+L.
$\label{enc2:forbidden3}$ :- forbidden(PAT,PER,ST,END), session(ID,PAT,_), extstart(ID,PER,TS), extlength(ID,PER,L), ST<=TS,END>TS.
$\label{enc2:fair}$ :- time(PER,_,T), macro_location(MAC,LOC1), macro_location(MAC,LOC2), #sum{1,ID1:session_time(ID1,_,LOC1,PER,T); -1,ID2:session_time(ID2,_,LOC2,PER,T)} > 2.
$\label{enc2:weakLength}$ :~ length(ID,_, L), session_length(ID,MIN,IDEAL), D=|L-IDEAL|. [D@6, ID]
$\label{enc2:weakPeriodPrioritized}$ :~ start(ID,PER,_), session_type(ID,_,individual), session_preference(ID,PER2,_,high), D=|PER-PER2|. [D@5, ID]
$\label{enc2:weakTimePrioritized}$ :~ start(ID,PER,TS), session_type(ID,_,individual), session_preference(ID,PER,TS2,high), D=|TS-TS2|. [D@4, ID]
$\label{enc2:weakOptionals}$ :~ optional_session(ID), time(PER,_,TS), not start(ID,PER,TS). [1@3,ID]
$\label{enc2:weakPeriodUnprioritized}$ :~ start(ID,PER,_), session_preference(ID,PER2,_,low), session_type(ID,_,individual), optional_session(ID), D=|PER-PER2|. [D@2, ID]
$\label{enc2:weakTimeUnprioritized}$ :~ start(ID,PER,TS), session_preference(ID,PER,TS2,low), session_type(ID,_,individual), optional_session(ID), D=|TS-TS2|. [D@1, ID]
    \end{asp}
    \caption{ASP Encoding for the agenda problem.}
    $\label{encoding:agenda}$
\end{figure*}

\paragraph{Encoding.} In Figure \ref{encoding:agenda} the encoding for the agenda is presented. Rules $r_{\ref{enc2:guessStartMandatory}}$ and $r_{\ref{enc2:guessStartOptional}}$ assign a start time to every session: for the optional session, the start atom can be unassigned.
Rule $r_{\ref{enc2:guessLength}}$ defines a length for all the sessions: the session length cannot be lower than the minimum time of the session and cannot be greater than the ideal time the session should take.
Rule $r_{\ref{enc2:guessSessionLocation}}$ assigns a location for each session.
Rules $r_{\ref{enc2:guessBefore}}$ and $r_{\ref{enc2:guessAfter}}$ reserve to each session slots of time before it starts and after it ends, in which the session can be performed in a supervised fashion. 

Then, rules $r_{\ref{enc2:extstart}}$ and $r_{\ref{enc2:extlength}}$ define auxiliary atoms \texttt{extstart(ID,PER,TS)} and \texttt{extlength(ID,PER,TS)} using \texttt{TS} slots of times for the session with identifier \texttt{ID} on period \texttt{PER} reserved for the start and length extensions, respectively.
Rule $r_{\ref{enc2:individualSessionLocation}}$ defines an auxiliary atom of the form \texttt{individual\_session\_location(ID,LOC,OP,MIN,IDEAL)} which represents that an individual session \texttt{ID} in the location \texttt{LOC} is assigned to the operator \texttt{OP}, and its minimum and ideal lengths are equal to \texttt{MIN} and \texttt{IDEAL}, respectively.
Rule $r_{\ref{enc2:sessionAtTime}}$ defines \texttt{session\_time(ID,OP,PL,PER,T)} which states that during time \texttt{T} of period \texttt{PER} the session \texttt{ID} is being performed by operator \texttt{OP}.

Rule $r_{\ref{enc2:individualAssignmentsOverlap}}$ states that two individual assignments shall not overlap.
Rule $r_{\ref{enc2:sessionPerPeriod}}$ imposes that each patient is assigned to at most one session per period.
Rules $r_{\ref{enc2:equalReduction1}}$ trough $r_{\ref{enc2:equalReduction3}}$ impose that the optional individual time (i.e., the difference between the minimum length of the session and the planned length) is added fairly to all individual sessions, starting with shorter ones.
Rule $r_{\ref{enc2:ubiquity}}$ imposes that for each time slot, the operator is not in two different places.
Rule $r_{\ref{enc2:minTimeReserved}}$ states that patients must have their minimum time reserved.
Rule $r_{\ref{enc2:gymCapacity}}$ imposes a limit on the concurrent use of locations with limited capacity.
Rules $r_{\ref{enc2:forbidden1}}$ through $r_{\ref{enc2:forbidden3}}$ impose that a session cannot happen during a forbidden time.
Rule $r_{\ref{enc2:fair}}$ avoids that, during a time slot, the distribution of sessions between each pair of locations inside the same macro location is unfair (i.e., a location is at its full capacity while another is empty).

The weak constraint $r_{\ref{enc2:weakLength}}$ states that each session duration should be as close as possible to the ideal duration.
Rules $r_{\ref{enc2:weakPeriodPrioritized}}$ and $r_{\ref{enc2:weakTimePrioritized}}$ minimize the distance between the actual and the preferred starting time for the  sessions with \textit{high} priority. 
Rule $r_{\ref{enc2:weakOptionals}}$ maximizes the number of optional sessions included in the scheduling.
Rules $r_{\ref{enc2:weakPeriodUnprioritized}}$ and $r_{\ref{enc2:weakTimeUnprioritized}}$ are similar to $r_{\ref{enc2:weakPeriodPrioritized}}$ and $r_{\ref{enc2:weakTimePrioritized}}$, respectively, but for the sessions having \textit{low} priority. 

\section{Experimental Analysis}
\label{sec:res}

In this section, the analyses performed on the two encodings is presented. The first part of our analysis is performed on real data 
coming from the institutes of Genova Nervi and Castel Goffredo; then, in order to evaluate the scalability of the approach and to analyse how our solution would behave in larger institutes having similar characteristics, an analysis is performed on synthetic instances with increasing dimensions, but considering real parameters. A comparison between the real and synthetic instances validates the approach and demonstrates that synthetic instances can reasonably model the problem at hand. All these three parts are included, in separate paragraphs, in a first subsection, while a second subsection is devoted to a comparison to alternative logic-based formalisms. Encodings and benchmarks used in the experiments can be found at:
\url{http://www.star.dist.unige.it/~marco/RuleMLRR2021TPLP/material.zip}.

\subsection{Results of the encoding}

\begin{table}[tb]
    \caption{Dimensions of the ICS Maugeri's institutes. 
    }
	\label{tab:hospitals}
	\centering
	\begin{tabular}{|c|c|c|c|c|c|}
	\hline
		Institute & \# Operators & \# Patients & Density & \# Floors & \# Gyms \\ \hline
		Genova Nervi & [9,18] & [37,67] & [2.4,5.2] & 1 & 1 \\\hline
		Castel Goffredo & [11,17] & [51,78] & [3.5, 6.4] & 2 & 3 \\
	\hline
	\end{tabular}
\end{table}

\begin{table}[tb]
\caption{Results on ICS Maugeri institutes. 
}
\label{tab:real-results}
\centering
\begin{tabular}{|c|c|c|c|c|c|c|c|c|}
\hline
\multicolumn{1}{|l|}{} & \multicolumn{4}{c|}{Branch \& Bound + RoM} & \multicolumn{4}{c|}{Unsatisfiable Core} \\ \hline
 & \multicolumn{2}{c|}{Genova Nervi} & \multicolumn{2}{c|}{Castel Goffredo} & \multicolumn{2}{c|}{Genova Nervi} & \multicolumn{2}{c|}{Castel Goffredo} \\ \hline
 & Board & Agenda & Board & Agenda & Board & Agenda & Board & Agenda \\ \hline
\% Optimum & 35\% & 0\% & 0\% & 0\% & 22\% & 45\% & 0\% & 0\% \\ \hline
\% Satisfiable & 65\% & 100\% & 100\% & 67\% & 78\% & 55\% & 100\% & 70\% \\ \hline
\% Unknown & 0\% & 0\% & 0\% & 33\% & 0\% & 0\% & 0\% & 30\% \\ \hline\hline
Avg Time for opt & 1.1s & - & - & - & 10s & 0.01s & - & - \\ \hline
Avg Time Last SM & 1.3s & 30s & 5.2s & 30s & 12.1s & 21.3s & 10.4s & 30s \\ \hline
\end{tabular}
\end{table}
\paragraph{Real data.}  ICS Maugeri utilizes a web-based software called QRehab \citep{PMID:33995247}, which is built on top of the specified encoding; thus, analysis can be performed on real data coming from the institutes of Genova Nervi and Castel Goffredo, which tested and used this software since mid 2020 for Genova Nervi and the beginning of 2021 for Castel Goffredo. This allowed us to access $290$ instances for Genova Nervi and $100$ for Castel Goffredo. Table \ref{tab:hospitals} provides an overview of the dimension of the instances in the two institutes in terms of number of physiotherapists, number of daily patients, density of patients per operator, number of floors (i.e., macro-locations) and number of total gyms (which we recall are locations in which multiple sessions can be performed in parallel). In Table \ref{tab:real-results}, the results obtained by the two encodings are presented in terms of percentage of instances for which an optimal/satisfiable/no solution is computed, which also correspond to the three outcomes of interest for a practical use of our solution. The last two rows report the mean time of instances solved optimally and of the last computed solution for all satisfiable instances, respectively. The scheduling was performed using the ASP solver {\sc clingo} \citep{DBLP:journals/ai/GebserKS12} with a cut-off of $30s$ using two different optimization methods: The first is the default Branch\&Bound (BB) optimization method \citep{DBLP:conf/lpnmr/GebserKK0S15} with the option \code{--restart-on-model} enabled; the second leverages instead the Unsatisfiable Core (USC) algorithm \citep{DBLP:conf/iclp/AndresKMS12} with the {\sc clingo} options \code{--opt-strategy=usc,k,0,4} and \code{--opt-usc-shrink=bin} enabled (which turn on the algorithm $k$~\citep{DBLP:journals/fuin/AlvianoD20} and the shrinking of the unsatisfiable cores \citep{DBLP:journals/tplp/AlvianoD16}, respectively). The cut-off of $30s$ was chosen in order to be able to analyse a vast amount of experiments in overall reasonable time, and has proven to be a sufficient amount of time to achieve meaningful results; in the software used daily by the ICS Maugeri the cut-off is set to $300s$, as a means to solve even the hardest instances, having a limited number of instances to be run daily. As it can be seen in Table \ref{tab:real-results}, results are mixed: the USC algorithm performs better in the agenda encoding while the BB algorithm is better on the board scheduling; moreover, 100\% of the board instances are solved, while for approximately one third of the agenda instances from Castel Goffredo a solution cannot be found. Considering these are hard real instances and cut-off time is limited, results are positive and highly appreciated by ICS Maugeri members.

\begin{figure}[thb]
\includegraphics[width=\textwidth]{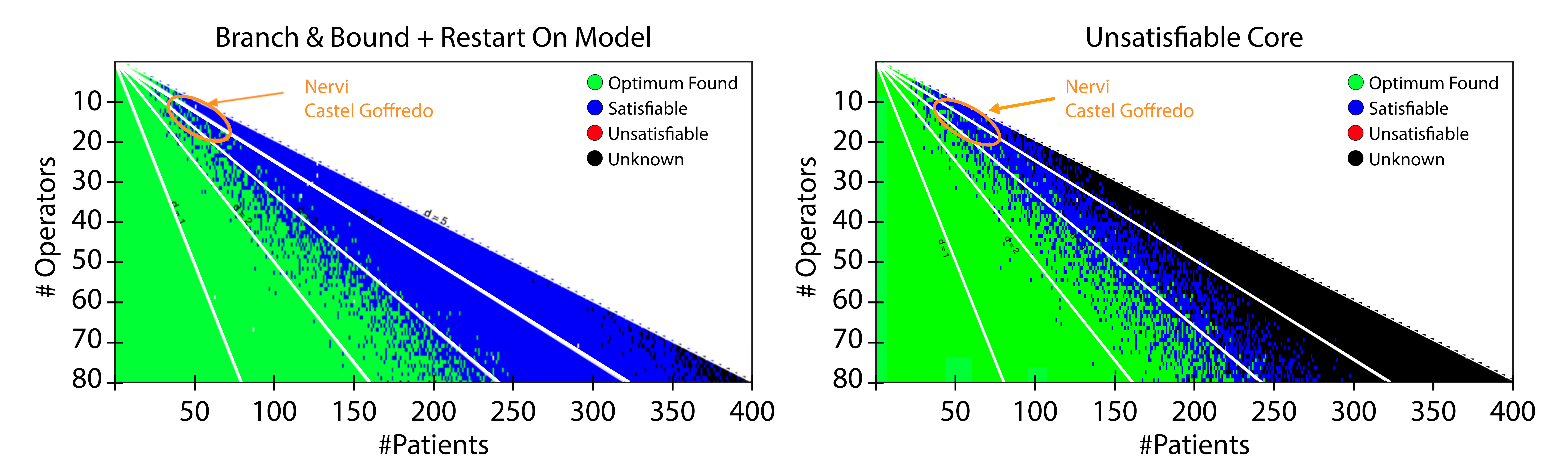}

\caption{Results of {\sc clingo} using the BB optimization algorithm  and the option \texttt{--restart-on-model} enabled (left) and the USC optimization algorithm (right) on synthetic benchmarks of the board. 
}\label{fig:rom-board}
\end{figure}

\paragraph{Synthetic data.} In order to understand how our solution scales to larger institutes having similar characteristics, a simulated approach is needed. For this reason, a generator able to produce random instances with features as close as possible to the ones of real hospitals was developed. Some examples of real data utilized are: the percentage of individual and supervised sessions, the medium length of operator's shifts, the occurrence of forbidden time slots for patients, and the ideal length of sessions. For every new instance created, each feature was extracted from a random distribution which was modelled from the real data coming from the hospitals or from the knowledge of institute administrators and managers. In Figure \ref{fig:rom-board} results of the scheduling of the board encoding, computed from the synthetic data, are presented. The x-axis defines the number of patients and the y-axis the number of operators; white lines represent points in which the density is an integer. Every pixel of the image depicts the mode of the results of 5 simulations executed with the corresponding number of patients and operators with a cut-off of $30s$ using the BB optimization algorithm (left) and the USC optimization algorithm (right). The colour of a pixel thus signals if the majority of instances with that particular number of operators and patients resulted in: (i) \textit{Optimum Found}, signalling that the optimal stable model was found, (ii) \textit{Satisfiable}, when at least one suboptimal stable model was found, but the solution is not guaranteed to be optimal, (iii) \textit{Unknown}, if no stable model could be found before the cut-off, or (iv) \textit{Unsatisfiable}, when no stable model exists which can satisfy all the constraints. As it can be seen from the figure, the results of the scheduling are directly proportional to the density (i.e., the average number of patients for every operator), changing from \textit{Optimum Found} to \textit{Satisfiable} when reaching a density of approximatively $2.4$ patients per operator. Notably, despite the use of random instances, no instance results \textit{Unsatisfiable} since the fictitious operator can always catch the patients which cannot be assigned to any operator (due to all the operators reaching full capacity). The position of the hospitals of Genova Nervi and Castel Goffredo are highlighted with a circle. In this figure it can be noted how BB gives better results than USC, by being able to find, before the cut-off, at least a suboptimal stable model for instances of higher densities, where, instead, the USC algorithm returns \textit{Unknown}. 

\begin{figure}[thb]
\includegraphics[width=\textwidth]{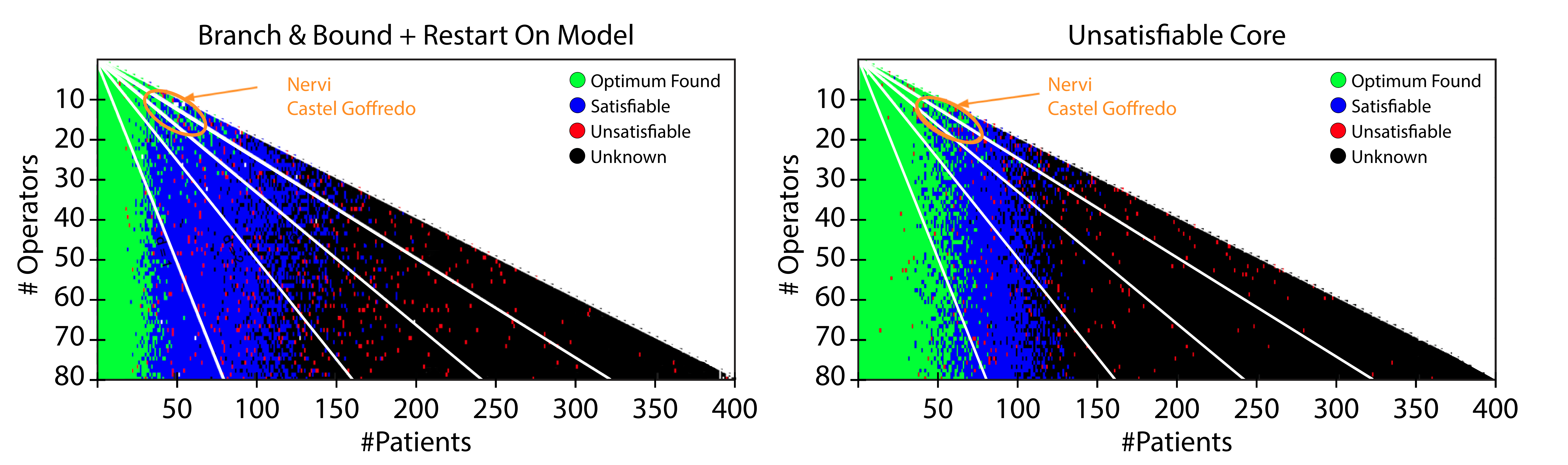}

\caption{Results of {\sc clingo} using the BB optimization algorithm  and the option \texttt{--restart-on-model} enabled (left) and the USC optimization algorithm (right) on synthetic benchmarks of the agenda.}\label{fig:rom-agenda}
\end{figure}

In Figure \ref{fig:rom-agenda} the results of the agenda encoding, scheduled with the BB algorithm (left) and USC algorithm (right), are presented in the same format as for the previous experiment. The instances for this experiment are the same as the previous one, but are augmented with the assignment among patients and operators found by {\sc clingo} with the board encoding and other needed parameters. As previously stated, each pixel represents 5 instances and its colour represents the mode of the \clingo results. Here two things can be noted: (i) unlike the board results, which showed a proportionality with the density, these results show a correlation only with the number of patients, and (ii) some red dots scattered in the image indicate that some instances result \textit{Unsatisfiable}: this can happen since the random data could create some instances with features that cause an impossibility to schedule. With the BB optimization algorithm, the transition between the \textit{Optimum Found} and \textit{Satisfiability} results is located near $40$ patients, and near $120$ patients for the transition between \textit{Satisfiability} and \textit{Unknown}. As it can be seen in Figure \ref{fig:rom-agenda} (right), the USC algorithm performs instead better and moves the transition between the \textit{Optimum found}  and \textit{Satisfiable} results from $40$ to $60$ patients but, on the other hand, the transition between \textit{Satisfiable} and \textit{Unknown} slightly decreases from $120$ patients to $110$. The improvements on the transition betwee \textit{Optimum found} and \textit{Satisfiable} is very important in our setting, since Genova Nervi and Castel Goffredo fall into this area, confirming the improvements obtained in Table \ref{tab:real-results}. 

\begin{figure}[thb]
\includegraphics[width=\textwidth]{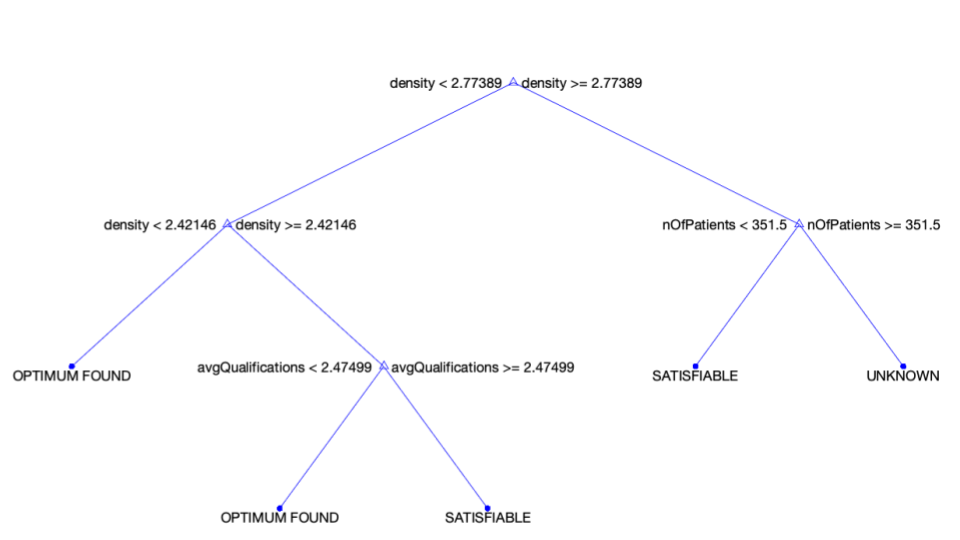}

\caption{Visual representation of the decision tree trained on the results found by \clingo on real data utilizing the BB+RoM algorithm. The tree nodes represent features of the instance (density and average qualifications) and the leafs represent the result given by \clingo (optimal found, satisfiable, unsatisfiable, unknown).}\label{fig:decision-tree}
\end{figure}

\paragraph{Validation of synthetic instances.} In order to understand if the simulated instances correctly represent the real data and can be therefore used to predict the behaviour of the system in larger institutes with similar characteristics, a validation is needed to compare the results obtained on real and synthetic instances. Intuitively, we have considered the data presented in Table \ref{tab:real-results} and compared it to the results of the instances within the circles around Genova Nervi and Castel Goffredo in Figures \ref{fig:rom-board} and \ref{fig:rom-agenda}, to check if they ``coincide". For doing so, a decision tree was trained, taking as dataset all the features of the simulated instances, some of them listed in the previous paragraph. Then, a test dataset with features extracted from the real instances was produced and given as input to the decision tree, and the predicted result was then compared to the result given by {\sc clingo} on the real instances. Figure \ref{fig:decision-tree} shows a visual representation of the decision tree trained on the results found by \clingo on real data utilizing the BB+RoM approach. The tree nodes represent the most important features found by the decision tree approach, which allows a correct classification of the results of \clingo, shown in the leaf nodes. The shown features are (i) the density, i.e., the proportion patients/operator ratio, and (ii) the average number of qualifications, i.e., the type of patients (orthopaedic, neurological, covid-positive etc). Synthetic resources were then used as new data and given as input to the decision tree. The output of the decision tree was then compared with the actual result given by \clingo. 
It can be seen, in fact, how the decision tree in Figure \ref{fig:decision-tree} is able to explain the results of the synthetic instances depicted in Figure \ref{fig:rom-board} (left): the colour green (representing optimality) is indeed classifiable only by the patients/operator density (i.e., the white diagonal lines) and the transition between the two results happen when the density is near 2.4, as classified, from the real data, by the decision tree. When the density is between 2.42 and 2.77 it can be noted how the average number of qualification (not explicitly shown in Figure \ref{fig:rom-board}) makes the difference between an optimal and suboptimal result (the more specialized the operators are, the fewer patients there are available for the planner to choose). This test showed that for the board encoding, all the results on real instances were equal to the predicted ones for both institutes; the agenda encoding produced instead the same results in $93\%$ of the cases for Genova Nervi and in $67\%$ of the cases for Castel Goffredo,
thus showing that, overall, the synthetic data behaves similarly to the real one and can be used for predicting the behavior of instances in larger institutes having similar characteristics. Finally, the computed decision trees also confirm what are the most relevant features outlined above by inspecting the figures. In fact, if the height of the decision tree is increased, the accuracy of prediction does not improve that much, signalling that the features shown in Figure \ref{fig:decision-tree} are sufficient to explain the differences in the results of \clingo.

\subsection{Comparison to alternative logic-based formalisms}
\label{subsec:comp}
In the following, an empirical comparison of our ASP-based solution to alternative logic-based approaches is presented, obtained by applying automatic translations of our ASP encoding.
In more detail, the ASP solver \textsc{wasp}~\citep{DBLP:conf/lpnmr/AlvianoADLMR19} was used, with the option \texttt{--pre=wbo}, which converts ground ASP instances into pseudo-Boolean instances in the wbo format~\citep{pbcompetition}.
Then, the tool \textsc{pypblib}~\citep{pypblib} was employed to encode wbo instances as MaxSAT instances.
Moreover, given that other formalisms cannot handle multi-level optimizations, in order to provide a fair comparison, the ASP instances were processed using \textsc{wasp} with the option \texttt{--pre=lparse}, which collapses all weak constraints levels into one single level using exponential weights. With this approach, the costs found by the different approaches can be straightforwardly compared.

Three state-of-the-art MaxSAT solvers were considered, namely \textsc{MaxHS}~\citep{DBLP:conf/sat/SaikkoBJ16}, \textsc{open-wbo}~\citep{DBLP:conf/sat/MartinsML14}, and \textsc{rc2}~\citep{DBLP:journals/jsat/IgnatievMM19}, as well as the industrial tool for solving optimization problems \textsc{gurobi}~\citep{gurobi}, which is able to process instances in the wbo format.
Concerning \textsc{clingo}, the already presented BB+RoM and USC algorithms were used. The latter enables the usage of algorithm \textsc{oll}~\citep{DBLP:conf/cp/MorgadoDM14}, which is the same algorithm employed by the MaxSAT solver \textsc{rc2}.

\begin{table}[tb]
\caption{Comparison between alternative logic-based formalisms for the board and agenda phase.}
\label{tab:asp2logic}
\centering
\resizebox{\textwidth}{!}{%
\begin{tabular}{cc|cccccc|c|ccccc|}
\cline{3-8} \cline{10-14}
 &  & \multicolumn{6}{c|}{Board} &  & \multicolumn{5}{c|}{Agenda} \\ \cline{3-8} \cline{10-14} 
 &  & \multicolumn{1}{c|}{BB+RoM} & \multicolumn{1}{c|}{USC} & \multicolumn{1}{c|}{MaxHS} & \multicolumn{1}{c|}{OpenWBO} & \multicolumn{1}{c|}{RC2} & Gurobi &  & \multicolumn{1}{c|}{USC} & \multicolumn{1}{c|}{MaxHS} & \multicolumn{1}{c|}{OpenWBO} & \multicolumn{1}{c|}{RC2} & Gurobi \\ \cline{1-1} \cline{3-8} \cline{10-14} 
\multicolumn{1}{|c|}{First} &  & \multicolumn{1}{c|}{58\%} & \multicolumn{1}{c|}{41,7\%} & \multicolumn{1}{c|}{0,3\%} & \multicolumn{1}{c|}{-} & \multicolumn{1}{c|}{-} & - &  & \multicolumn{1}{c|}{77,1\%} & \multicolumn{1}{c|}{15,6\%} & \multicolumn{1}{c|}{7,3\%} & \multicolumn{1}{c|}{-} & - \\ \cline{1-1} \cline{3-8} \cline{10-14} 
\multicolumn{1}{|c|}{Second} &  & \multicolumn{1}{c|}{41,7\%} & \multicolumn{1}{c|}{58,3\%} & \multicolumn{1}{c|}{1\%} & \multicolumn{1}{c|}{-} & \multicolumn{1}{c|}{-} & - &  & \multicolumn{1}{c|}{14,7\%} & \multicolumn{1}{c|}{54,2\%} & \multicolumn{1}{c|}{29,4\%} & \multicolumn{1}{c|}{-} & - \\ \cline{1-1} \cline{3-8} \cline{10-14} 
\multicolumn{1}{|c|}{Third} &  & \multicolumn{1}{c|}{0,3\%} & \multicolumn{1}{c|}{-} & \multicolumn{1}{c|}{10,3\%} & \multicolumn{1}{c|}{-} & \multicolumn{1}{c|}{-} & - &  & \multicolumn{1}{c|}{6,4\%} & \multicolumn{1}{c|}{28,4\%} & \multicolumn{1}{c|}{61,5\%} & \multicolumn{1}{c|}{-} & - \\ \cline{1-1} \cline{3-8} \cline{10-14} 
\multicolumn{1}{|c|}{Solver TO} &  & \multicolumn{1}{c|}{-} & \multicolumn{1}{c|}{-} & \multicolumn{1}{c|}{9,3\%} & \multicolumn{1}{c|}{20,9\%} & \multicolumn{1}{c|}{20,9\%} & 100\% &  & \multicolumn{1}{c|}{1,8\%} & \multicolumn{1}{c|}{1,8\%} & \multicolumn{1}{c|}{1,8\%} & \multicolumn{1}{c|}{100\%} & 100\% \\ \cline{1-1} \cline{3-8} \cline{10-14} 
\multicolumn{1}{|c|}{Pypblip TO} &  & \multicolumn{1}{c|}{-} & \multicolumn{1}{c|}{-} & \multicolumn{1}{c|}{79,1\%} & \multicolumn{1}{c|}{79,1\%} & \multicolumn{1}{c|}{79,1\%} & - &  & \multicolumn{1}{c|}{-} & \multicolumn{1}{c|}{-} & \multicolumn{1}{c|}{-} & \multicolumn{1}{c|}{-} & - \\ \cline{1-1} \cline{3-8} \cline{10-14} 
\end{tabular}%
}
\end{table}

These experiments were run using the ASP encoding coming from the already presented real-world instances of the hospitals of Genova Nervi and Castel Goffredo. The experiments were conducted in the following way: firstly, the ASP encoding in which the weak constraints have been collapsed, was transformed in the wbo and MaxSAT formats, then all the solvers were called with a cut-off of $30s$ (the same used in all the other experiments). Then, for every formalism the following metrics were recorded: if it has found the optimum, the final cost and the time of computation. With these metrics, the formalisms can be ordered from best to worst based on their result: an optimal solution is better than one not declared optimal; if both are suboptimal then the one with the lowest cost is better; if both are optimal then the one which took less to declare optimality is better. In Table \ref{tab:asp2logic} the ranking among the formalisms is presented. For each of the different formalisms, the table shows the percentage of how many times it has arrived first, second or third. \textit{Solver TO} represents cases in which the solver was unable to find a solution before the cut-off (\textit{Unknown}). \textit{Pypblib TO} represents cases in which the tool \textsc{pypblib} was unable to encode wbo instances as MaxSAT instances in a cut-off time of $60s$. For the board encoding both \textsc{clingo}'s algorithms USC and BB+RoM are presented, since they showed comparable result;  instead, in the agenda encoding only the USC algorithm is presented since it outperformed the BB+RoM in the previous tests.

The results show that for the board encoding \textsc{clingo} is the most performant algorithm, coming first in almost all the experiments. In particular, \textsc{clingo} with the BB+RoM optimization algorithm resulted more performant than the algorithm relying on the Unsatisfiable Core strategy, which is conformant with the experiments run on the multi-level version of the ASP encoding. For this encoding, it can be seen that for a high number of instances, around 80\%, the tool \textsc{pypblib} was unable to encode the MaxSAT instances within the cut-off. Still, in the remaining 20\%, \textsc{clingo} remains the most performant algorithm. For the agenda encoding, \textsc{clingo} is still the best solver, but a more precise ranking among solvers can be noticed with MaxHS coming second and OpenWBO third. Notably, RC2 and Gurobi are, with both encoding, always unable to find a solution within the cut-off.

\section{Domain Specific Optimisations}
\label{sec:opt}
Motivated by the analysis performed in the previous section, in which ASP outperformed other formalisms on translated (MaxSAT and pseudo-Boolean) formulas, we apply domain specific optimizations to our ASP encoding, with the aim of further improving the solving time and move towards solving larger instances. In Section \ref{sec:res}, benchmarks for the board and agenda phases were presented, showing different results based on the optimization algorithm chosen (i.e., BB+RoM or USC). The domain specific optimisations are presented to mainly decrease grounding,  and consequently planning times, with the aim, as mentioned, of being able to find optimum solutions in larger instances, e.g., possibly corresponding to larger hospitals. The optimizations are presented only on the agenda encoding, which is the more complicated of the two phases and has still a great margin of improvement via changes in the encoding. These changes all rely on the knowledge of the RSP domain and on the possibility to prune impossible solutions already in the grounding process, avoiding wasting time in search. The section is organized in two subsections, in which the first presents the changes and improvements done on the previous agenda encoding, while the second subsection focuses on the results.

\subsection{Optimized encoding}
The next two paragraphs present the specific domain optimizations introduced. 
\paragraph{Pruning of session starts.} As it can be seen in Figure \ref{encoding:agenda}, in rules $r_{\ref{enc2:guessStartMandatory}}$ and $r_{\ref{enc2:guessStartOptional}}$ the start of a session is guessed between all the possible time slots in the shift of an operator, expressed via the atom \texttt{time(PER,OP,TS)}. These guess rules can be improved by reducing the number of time slots in which it is possible to start a session with the following constraints: 
\begin{enumerate}
    \item a session cannot start in a time slot near to the operator' shift's end. This is because the minimum specified time of the session would not be satisfied, given the shift's end; 
    \item if a patient has a forbidden time (i.e., a time interval where the patient cannot be scheduled), the session cannot start during the forbidden time. Moreover, some timeslots before the forbidden times should be excluded beforehand since, if the session started in these timeslots, this would not allow it to end before the forbidden time starts.
\end{enumerate}

\begin{figure}[t!]
    \centering
    \begin{asp}[firstnumber=last]
$\label{enc:optstart:forbiddenRange}$forbiddenRange(ID,PER,XSTA,END) :- forbidden(PAT,PER,STA,END), session(ID,PAT,_), session_length(ID,MIN,_), XSTA = STA - MIN + 1.
$\label{enc:optstart:forbiddenSlot}$forbiddenSlot(ID,PER,STA..END-1) :- forbiddenRange(ID,PER,STA,END).
$\label{enc:optstart:allowedTime}$allowedTime(ID,PER,T) :- time(PER,OP,T), session(ID,_,OP), session_length(ID,MIN,_), period(PER,OP,_,END), T <= END - MIN, not forbiddenSlot(ID,PER,T).
$\label{enc:optstart:startMandatory}$1 {start(ID,PER,TS) : allowedTime(ID,PER,TS)} 1 :- session(ID,_,OP), mandatory_session(ID).
$\label{enc:optstart:startOptional}$0 {start(ID,PER,TS) : allowedTime(ID,PER,TS)} 1 :- session(ID,_,OP), optional_session(ID).
\end{asp}
\caption{Optimized encoding for pruning the session starts}
\label{encoding:optstart}
\end{figure}

In Figure \ref{encoding:optstart} the ASP encoding for pruning the session starts is shown. In $r_{\ref{enc:optstart:forbiddenRange}}$ a new atom \texttt{forbiddenRange} is defined for the purpose of extending the start of forbidden times by including the time slots which would not allow the session to end before the start of the forbidden time. Rule $r_{\ref{enc:optstart:forbiddenSlot}}$ spreads  \texttt{forbiddenRange} in all the time slots (\texttt{forbiddenSlot}) within the range. In rule $r_{\ref{enc:optstart:allowedTime}}$ a new atom \texttt{allowedTime} is defined as a time slot in the shift of the operator which is not a \texttt{forbiddenSlot}, and which allows the session, with its min length \texttt{MIN}, to end before the end of the shift \texttt{END}. In rules $r_{\ref{enc:optstart:startMandatory}}$ and $r_{\ref{enc:optstart:startOptional}}$, the atom \texttt{allowedTime} replaces the more broad atom \texttt{time} in the guess rule of the start of the session. In the optimized encoding, rules from $r_{\ref{enc:optstart:forbiddenRange}}$ to $r_{\ref{enc:optstart:startOptional}}$ replace rules $r_{\ref{enc2:guessStartMandatory}}$ and $r_{\ref{enc2:guessStartOptional}}$ of the original encoding (Figure \ref{encoding:agenda}).

\paragraph{Pruning of session extension.} As stated in Section \ref{subsec:agenda}, the agenda encoding relies on two auxiliary atoms (\texttt{extstart} and \texttt{extlength}) as a means to reserve slots of time before it starts and after it ends to each session, in which the session can be performed in a supervised fashion. These before (after) slots of time are decided with a guess rule on the atom \texttt{before} (\texttt{after}) in rule $r_{\ref{enc2:guessBefore}}$ ($r_{\ref{enc2:guessAfter}}$) of Figure \ref{encoding:agenda}. The definition of these atoms can be improved in order to reduce the number of ground instantiations, in the following way:
\begin{enumerate}
    \item as described in the previous paragraph, the extended part of a session cannot start during a forbidden time;
    \item the \texttt{before} and \texttt{after} timeslots cannot be greater than the difference between the ideal length of a session and its minimum length. Since the weak constraints impose a minimization of the distance between the final length of the session and its ideal length, this last acts as an upper bound of the length of the session;
    \item if there is already an extension before the session, the extended length of the session cannot be longer than the ideal length of the session.
\end{enumerate}

\begin{figure}[t!]
    \centering
    \begin{asp}[firstnumber=last]
$\label{enc:optext:before}$1 {before(ID,NL): allowedTime(ID,PER,T), T<=TS, NL=TS-T, NL<=IDEAL-MIN} 1 :- start(ID,PER,TS), session(ID,_,OP), session_length(ID,MIN,IDEAL).
$\label{enc:optext:extstart}$extstart(ID,PER,TS-LB) :- start(ID,PER,TS), before(ID,LB).
$\label{enc:optext:after}$1 {after(ID,NL): time(PER,OP,T), T>=TS+LEN, NL=T-TS-LEN, NL<=IDEAL-MIN} 1 :- start(ID,PER,TS), length(ID,PER,LEN), session(ID,_,OP), session_length(ID,MIN,IDEAL).
$\label{enc:optext:extlength}$extlength(ID,PER,LEN) :- length(ID,PER,L), after(ID,LA), before(ID,LB), session(ID,_,_), session_length(ID,_,IDEAL), LEN=L+LA+LB, LEN <= IDEAL.
$\label{enc:optext:notextlength}$:- start(ID, _, _), not extlength(ID, _, _).

\end{asp}
\caption{Optimized encoding for pruning of session extension}
\label{encoding:optext}
\end{figure}

Rules $r_{\ref{enc2:guessBefore}}$, $r_{\ref{enc2:guessAfter}}$, $r_{\ref{enc2:extstart}}$ and $r_{\ref{enc2:extlength}}$ of the agenda encoding presented in Figure \ref{encoding:agenda} can be substituted by the encoding presented in Figure \ref{encoding:optext}. Rule $r_{\ref{enc:optext:before}}$ states that a value for the before extension can be computed taking the difference between the start of the session and an allowed time slot distant no more than the difference between the ideal and minimum length of a session. Rule $r_{\ref{enc:optext:extstart}}$ defines the auxiliary atom \texttt{extstart} via the previously guessed \texttt{before} atom. Rule  $r_{\ref{enc:optext:after}}$ finds the amount to reserve after the session in the same way as expressed with the \texttt{before} atom. Rule $r_{\ref{enc:optext:extlength}}$ defines the auxiliary atom \texttt{extlength} now being limited by the ideal length. Rule $r_{\ref{enc:optext:notextlength}}$ imposes that a session must have an extended length (which can correspond to the individual length if no supervised time is needed); this is to avoid solutions in which the planner increases to the maximum both the before and after extensions of a session as a shortcut to falsify all instantiations of rule $r_{\ref{enc:optext:extlength}}$ by not having an extended length less than the ideal length.

\subsection{Results of the optimized encoding}
The next two paragraphs present the performance of the optimized encoding on real and synthetic instances, respectively.

\begin{table}[tb]
\caption{Comparison, in terms of grounding, between the basic and the optimized encoding on real instances coming from the Maugeri's hospitals.}
\label{tab:grounding-times}
\centering
\begin{tabular}{|c|cc|cc|}
\hline
 & \multicolumn{2}{c|}{Basic} & \multicolumn{2}{c|}{Optimized} \\ \hline
 & \multicolumn{1}{c|}{Nervi} & C.G. & \multicolumn{1}{c|}{Nervi} & C.G. \\ \hline
Avg Grounding Time & \multicolumn{1}{c|}{8.3s} & 11.5s & \multicolumn{1}{c|}{0.7s} & 0.9s \\ \hline
Avg Number of Variables & \multicolumn{1}{c|}{323k} & 587k & \multicolumn{1}{c|}{51k} & 59k \\ \hline
Avg Number of Rules & \multicolumn{1}{c|}{3.8M} & 11.4M & \multicolumn{1}{c|}{177k} & 327k \\ \hline
\end{tabular}

\end{table}

\begin{table}[tb]
\caption{Results of the optimized agenda encoding on real instances}
\label{tab:real-results-opt}
\centering
\resizebox{\textwidth}{!}{%
\begin{tabular}{|c|cc|cc|cc|cc|}
\hline
 & \multicolumn{2}{c|}{Basic+BB+RoM} & \multicolumn{2}{c|}{Basic+USC} & \multicolumn{2}{c|}{Opt+BB+RoM} & \multicolumn{2}{c|}{Opt+USC} \\ \hline
 & \multicolumn{1}{c|}{Nervi} & C.G. & \multicolumn{1}{c|}{Nervi} & C.G & \multicolumn{1}{c|}{Nervi} & C.G. & \multicolumn{1}{c|}{Nervi} & C.G. \\ \hline
\% Optimum & \multicolumn{1}{c|}{0\%} & 0\% & \multicolumn{1}{c|}{45\%} & 0\% & \multicolumn{1}{c|}{0\%} & 0\% & \multicolumn{1}{c|}{82\%} & 74\% \\ \hline
\% Satisfiable & \multicolumn{1}{c|}{100\%} & 67\% & \multicolumn{1}{c|}{55\%} & 70\% & \multicolumn{1}{c|}{100\%} & 100\% & \multicolumn{1}{c|}{18\%} & 26\% \\ \hline
\% Unknown & \multicolumn{1}{c|}{0\%} & 33\% & \multicolumn{1}{c|}{0\%} & 30\% & \multicolumn{1}{c|}{0\%} & 0\% & \multicolumn{1}{c|}{0\%} & 0\% \\ \hline
Avg Time for opt & \multicolumn{1}{c|}{-} & - & \multicolumn{1}{c|}{0.01s} & - & \multicolumn{1}{c|}{-} & - & \multicolumn{1}{c|}{0.01s} & 4.7s \\ \hline
Avg Time Last SM & \multicolumn{1}{c|}{30s} & 30s & \multicolumn{1}{c|}{21.3s} & 30s & \multicolumn{1}{c|}{19.6s} & 20s & \multicolumn{1}{c|}{20.4s} & 8.0s \\ \hline
\end{tabular}%
}
\end{table}

\paragraph{Real Data.} In Table \ref{tab:grounding-times} and  \ref{tab:real-results-opt} the results of the optimized encoding on real instances of the hospitals in Genova Nervi and Castel Goffredo are presented. Table \ref{tab:grounding-times} presents a comparison about the grounding between the two encodings, showing the significant reduction in terms of time, number of variables and number of rules that the optimized encoding brings. Table \ref{tab:real-results-opt} then shows the results for the basic agenda encoding presented in Section \ref{sec:enc-agenda} with the two algorithms BB+RoM and USC (this part of the table is copied from Table \ref{tab:real-results}). The other half of the table shows the results for the optimized encoding presented in the previous subsection, again with the two algorithms. As it can be seen, the optimized encoding boosts the performances, especially when combined with the USC algorithm. Comparing the two encodings, the first thing that can be noticed is that the optimized  encoding is able to find a solution for each instance. Moreover, it can be seen that: 
\begin{itemize}
    \item even if in Genova Nervi the percentages remain the same for the BB algorithm, the average time in which the last stable model is outputted decreases. 
    \item with the USC algorithm, using the optimized encoding, for most of the instances both in Genova Nervi and Castel Goffredo, an optimal solution can be found.
\end{itemize}

\paragraph{Synthetic Data.}

\begin{figure}[thb]
\includegraphics[width=\textwidth]{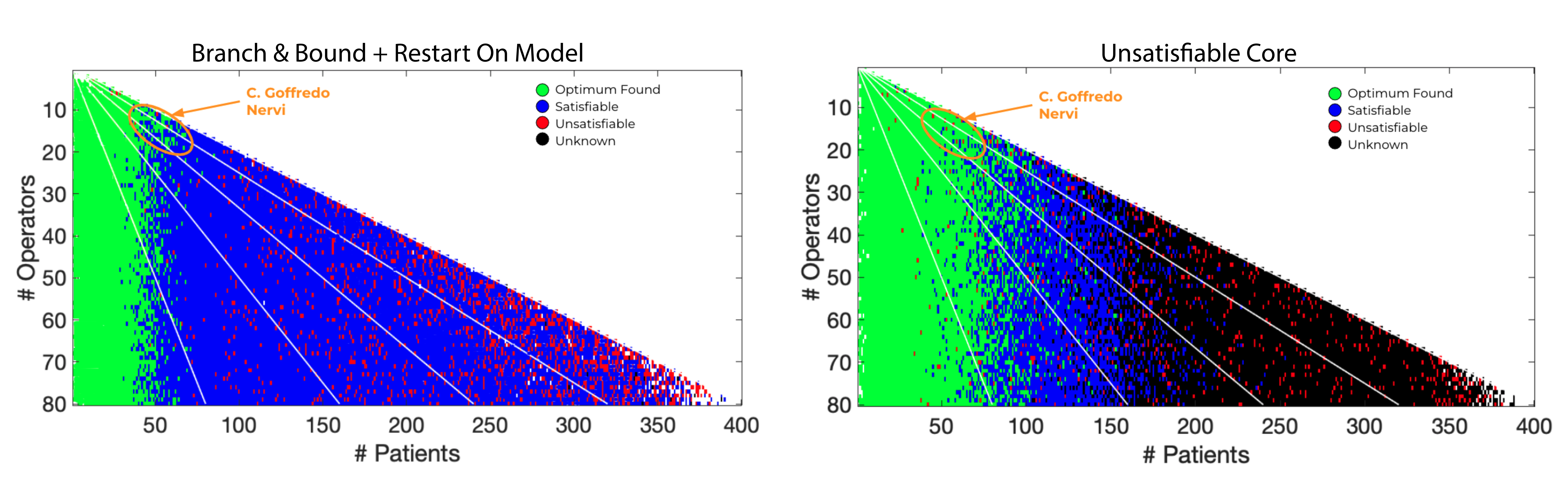}

\caption{Results of the synthetic benchmarks of the agenda produced by {\sc clingo} with the optimized encoding}\label{fig:opt-usc-rom-agenda}
\end{figure}

As previously explained in Section \ref{sec:res}, testing an encoding on synthetically generated instances is important to understand how our solution could scale to larger institutes having similar characteristics. Figure \ref{fig:opt-usc-rom-agenda} shows the results of the scalability test performed with the new optimized encoding, where the meaning of the colours, axes and lines has been already explained in Section \ref{sec:res}. On the left, the results of running the optimized encoding using \textsc{clingo}  with BB+RoM settings; on the right, the result are computing against the optimized encoding leveraging the USC algorithm. Comparing Figure \ref{fig:opt-usc-rom-agenda} (optimized encoding, Opt) with Figure \ref{fig:rom-agenda} (basic encoding, Basic) serious improvements can be noted: 
\begin{itemize}
    \item comparing the best combinations, i.e., Basic+USC and Opt+BB+RoM, it can be noted how the transition between solution with \textit{Optimum Found} and \textit{Satisfiable} stays approximatively the same near $50$ patients, but \textit{Unknown} results no longer appear, meaning that in the cut-off of $30s$ \textsc{clingo} can find a suboptimal solution. In fact, the aim of the optimization was to reduce to the minimum the grounding time, which has left now the largest part of the $30s$ cut-off to be spent in actually solving.
    \item focusing on Opt, the results of Opt+BB+RoM and Opt+USC show the supremacy of the USC algorithm.
\end{itemize}

Focusing on optimization algorithms, as it can be seen from the figures, the results on these benchmarks are comparable with the ones performed on real data: using the BB+RoM algorithm in fact it can be noted how the area of the graph with properties similar to the hospitals of Nervi and Castel Goffredo (the orange circle) have most of its area of a blue colour (representing Satisfiable results) and only for easy examples an optimal solution can be found; using the USC approach, similarly, we can see that in the circle fall most solutions found with optimality, thus confirming the results on the real instances. In fact, comparing the results of Basic+USC and Opt+USC it can be seen a real improvement in the number of instances which can be now solved with optimality: before, with the Basic+USC approach, the transition between solutions \textit{Optimum Found} and \textit{Satisfiable} lied near $50$ patients, while now has reached almost $90$ patients. 

\begin{table}[]
\centering
\caption{Comparison between alternative logic-based formalisms for the optimized agenda phase.}
\label{tab:asp2logic-opt}
\begin{tabular}{cc|ccccc|}
\cline{3-7}
 &  & \multicolumn{5}{c|}{Agenda} \\ \cline{3-7} 
 &  & \multicolumn{1}{c|}{USC} & \multicolumn{1}{c|}{MaxHS} & \multicolumn{1}{c|}{OpenWBO} & \multicolumn{1}{c|}{RC2} & Gurobi \\ \cline{1-1} \cline{3-7} 
\multicolumn{1}{|c|}{First} &  & \multicolumn{1}{c|}{94,2\%} & \multicolumn{1}{c|}{3,1\%} & \multicolumn{1}{c|}{0,0\%} & \multicolumn{1}{c|}{-} & - \\ \cline{1-1} \cline{3-7} 
\multicolumn{1}{|c|}{Second} &  & \multicolumn{1}{c|}{3,1\%} & \multicolumn{1}{c|}{72,3\%} & \multicolumn{1}{c|}{21,5\%} & \multicolumn{1}{c|}{-} & - \\ \cline{1-1} \cline{3-7} 
\multicolumn{1}{|c|}{Third} &  & \multicolumn{1}{c|}{0,0\%} & \multicolumn{1}{c|}{21,5\%} & \multicolumn{1}{c|}{75,5\%} & \multicolumn{1}{c|}{-} & - \\ \cline{1-1} \cline{3-7} 
\multicolumn{1}{|c|}{Solver TO} &  & \multicolumn{1}{c|}{2,7\%} & \multicolumn{1}{c|}{3,1\%} & \multicolumn{1}{c|}{3,0\%} & \multicolumn{1}{c|}{100\%} & 100\% \\ \cline{1-1} \cline{3-7} 
\multicolumn{1}{|c|}{Pypblip TO} &  & \multicolumn{1}{c|}{-} & \multicolumn{1}{c|}{-} & \multicolumn{1}{c|}{-} & \multicolumn{1}{c|}{-} & - \\ \cline{1-1} \cline{3-7} 
\end{tabular}
\end{table}

At last, we also compared our approach with algorithm USC (that the analysis demonstrated to be the best) to the other logic-based formalisms already employed in Table~\ref{tab:asp2logic}, using the same evaluation metric and presentation. As it is clear from Table~\ref{tab:asp2logic-opt}, the ASP approach is the best also when considering the optimized encoding.

\section{Related Work}
\label{sec:rel}

This paper is an extended and revised version of \citep{DBLP:conf/ruleml/CardelliniNDGGM21}, having the following main consistent additions: $(i)$ a comparison to alternative logic-based formalisms on real instances (Subsection \ref{subsec:comp}), and $(ii)$ the definition and related experimental evaluation of two domain specific optimizations (Section \ref{sec:opt}).

There have been few attempts in the literature to solve  rehabilitation scheduling, since most hospitals are still doing it in a manual way. Among the automated solutions, often they are applied to real world data. However, their results are not directly comparable to ours, since their constraints and objective functions are different from the ones that emerged from our meetings with the physiotherapists and management at ICS Maugeri. In particular, to our knowledge, no other solution takes into account several aspects like the preferred time for the session scheduling and the preferences in the assignment of the patient to the operator.  \citeauthor{Huang2012DecisionSS}~(\citeyear{Huang2012DecisionSS}) developed a system, equipped with a Graphical User Interface, which can generate the optimal schedules for rehabilitation patients to minimize waiting time and thus enhance service quality and overall resource effectiveness of rehabilitation facilities. More recently, \citeauthor{huynh_hybrid_2018}~(\citeyear{huynh_hybrid_2018}) further refined the algorithm in order to develop a hybrid genetic algorithm (GASA) that integrates genetic algorithm (GA) and simulated annealing (SA). 
Recently, \citeauthor{Li_2021}~(\citeyear{Li_2021}) designed a genetic algorithm based on Waiting Time Priority Algorithm (WTPA), which was tested on a rehabilitation department. \citeauthor{schimmelpfeng_decision_2012}~(\citeyear{schimmelpfeng_decision_2012}) developed a decision support system for the scheduling process based on mixed-integer linear programs (MILPs), to determine appointments for patients of rehabilitation hospitals, subject to numerous constraints that are often found in practice. We already mentioned in the introduction that ASP has been already successfully used for solving application problems in several research areas (see, e.g., \cite{DBLP:conf/iclp/GebserKKOSW16} for routing driverless transport vehicles, \cite{DBLP:journals/tplp/RiccaGAMLIL12} for team scheduling, and \cite{DBLP:journals/aim/ErdemGL16} for a general overview) including scheduling problems in the Healthcare domain (see, e.g., \cite{DBLP:conf/aiia/AlvianoBCDGKMMM20} for an overview focused on them). Differently from previous papers in the Healthcare domain, the current work focuses on the rehabilitation scheduling problem, that was not addressed before using ASP; and it combines a two-phase encoding, rather than the usually employed direct encoding, with the evaluation of the solution on real benchmarks.
Concerning the experimental analysis, similarly to \cite{DBLP:journals/tplp/DodaroGGMMP21}, in this work we compared our ASP-based solution with alternative logic-based approaches.

Finally, in \citep{PMID:33995247} an analysis of the usage of the tool in the hospital of Genova Nervi for a period of approx. 6 months is reported. As an example, statistics about the sessions planned by our ASP encodings and their actual durations in the hospital usage are recorded. As shown in the paper, reported and planned session lengths are similar, with the ratio between these two quantities has been between 0.95 and 1.05 for the 95\% of the considered time span. 


\section{Conclusion}
\label{sec:conc}

In this paper, we have presented a two-phase ASP encoding for solving rehabilitation scheduling. We have started from a general solution, then extended with domain specific optimizations. Our solution has been tested with {\sc clingo} and both real and synthetic benchmarks, the former provided by ICS Maugeri while the latter created with real parameters and employed to understand a possible behavior of the solution on upcoming institutes where the solution will be employed. Results are satisfying for the institutes employed at the moment and give some indications about the upcoming institutes ICS Maugeri plans to instrument with this solution.
Domain specific optimizations further improve the results, by also diminishing the number of instances for which a solution cannot be found in short time. Future research includes a more fine-grained analysis of our solution by, e.g., combining the strengths of the optimization algorithms, analysing further dimensions of our encoding, e.g., number of floors and gyms for synthetic benchmarks, and benchmarking the impact of the introduced domain specific optimizations separately.\\  

\noindent
\textbf{Competing interests:} The authors declare none.

\bibliographystyle{tlplike}
\bibliography{bibtex}

\begin{thebibliography}{}

\bibitem[Alviano et~al., 2019]{DBLP:conf/lpnmr/AlvianoADLMR19}
{\sc Alviano, M.}, {\sc Amendola, G.}, {\sc Dodaro, C.}, {\sc Leone, N.}, {\sc
  Maratea, M.}, {\sc and} {\sc Ricca, F.}
\newblock Evaluation of disjunctive programs in {WASP}.
\newblock In {\em {LPNMR} 2019} 2019, volume 11481 of {\em LNCS}, pp. 241--255.
  Springer.

\bibitem[Alviano et~al., 2020]{DBLP:conf/aiia/AlvianoBCDGKMMM20}
{\sc Alviano, M.}, {\sc Bertolucci, R.}, {\sc Cardellini, M.}, {\sc Dodaro,
  C.}, {\sc Galat{\`{a}}, G.}, {\sc Khan, M.~K.}, {\sc Maratea, M.}, {\sc
  Mochi, M.}, {\sc Morozan, V.}, {\sc Porro, I.}, {\sc and} {\sc Schouten, M.}
\newblock Answer set programming in healthcare: Extended overview.
\newblock In {\em Joint Proceedings of the 8th IPS Workshop and the 27th RCRA
  Workshop co-located with AIxIA 2020} 2020, volume 2745 of {\em {CEUR}
  Workshop Proceedings}. CEUR-WS.org.

\bibitem[Alviano and Dodaro, 2016]{DBLP:journals/tplp/AlvianoD16}
{\sc Alviano, M.} {\sc and} {\sc Dodaro, C.} 2016.
\newblock Anytime answer set optimization via unsatisfiable core shrinking.
\newblock {\em Theory and Practice of Logic Programming},  {\it 16}, 5-6,
  533--551.

\bibitem[Alviano and Dodaro, 2020]{DBLP:journals/fuin/AlvianoD20}
{\sc Alviano, M.} {\sc and} {\sc Dodaro, C.} 2020.
\newblock Unsatisfiable core analysis and aggregates for optimum stable model
  search.
\newblock {\em Fundamenta Informaticae},  {\it 176}, 3-4, 271--297.

\bibitem[Andres et~al., 2012]{DBLP:conf/iclp/AndresKMS12}
{\sc Andres, B.}, {\sc Kaufmann, B.}, {\sc Matheis, O.}, {\sc and} {\sc Schaub,
  T.}
\newblock Unsatisfiability-based optimization in clasp.
\newblock In {\em Technical Communications of the 28th International Conference
  on Logic Programming, {ICLP} 2012} 2012, volume~17 of {\em LIPIcs}, pp.
  211--221. Schloss Dagstuhl - Leibniz-Zentrum fuer Informatik.

\bibitem[Ansótegui et~al., 2019]{pypblib}
{\sc Ansótegui, C.}, {\sc Pacheco, T.}, {\sc and} {\sc Pon, J.} 2019.
\newblock Pypblib.

\bibitem[Baral, 2003]{baral2003}
{\sc Baral, C.} 2003.
\newblock {\em Knowledge Representation, Reasoning and Declarative Problem
  Solving}.
\newblock Cambridge University Press.

\bibitem[Brewka et~al., 2011]{DBLP:journals/cacm/BrewkaET11}
{\sc Brewka, G.}, {\sc Eiter, T.}, {\sc and} {\sc Truszczynski, M.} 2011.
\newblock Answer set programming at a glance.
\newblock {\em Communications of the {ACM}},  {\it 54}, 12, 92--103.

\bibitem[Calimeri et~al., 2020]{DBLP:journals/tplp/CalimeriFGIKKLM20}
{\sc Calimeri, F.}, {\sc Faber, W.}, {\sc Gebser, M.}, {\sc Ianni, G.}, {\sc
  Kaminski, R.}, {\sc Krennwallner, T.}, {\sc Leone, N.}, {\sc Maratea, M.},
  {\sc Ricca, F.}, {\sc and} {\sc Schaub, T.} 2020.
\newblock {ASP}-{C}ore-2 input language format.
\newblock {\em Theory and Practice of Logic Programming},  {\it 20}, 2,
  294--309.

\bibitem[Cardellini et~al., 2021]{DBLP:conf/ruleml/CardelliniNDGGM21}
{\sc Cardellini, M.}, {\sc Nardi, P.~D.}, {\sc Dodaro, C.}, {\sc Galat{\`{a}},
  G.}, {\sc Giardini, A.}, {\sc Maratea, M.}, {\sc and} {\sc Porro, I.}
\newblock A two-phase {ASP} encoding for solving rehabilitation scheduling.
\newblock In {\sc Moschoyiannis, S.}, {\sc Pe{\~{n}}aloza, R.}, {\sc
  Vanthienen, J.}, {\sc Soylu, A.}, {\sc and} {\sc Roman, D.}, editors, {\em
  Proceedings of the 5th International Joint Conference on Rules and Reasoning
  (RuleML+RR 2021)} 2021, volume 12851 of {\em Lecture Notes in Computer
  Science}, pp. 111--125. Springer.

\bibitem[Cieza et~al., 2020]{cieza_global_2020}
{\sc Cieza, A.}, {\sc Causey, K.}, {\sc Kamenov, K.}, {\sc Hanson, S.~W.}, {\sc
  Chatterji, S.}, {\sc and} {\sc Vos, T.} 2020.
\newblock Global estimates of the need for rehabilitation based on the {Global}
  {Burden} of {Disease} study 2019: a systematic analysis for the {Global}
  {Burden} of {Disease} {Study} 2019.
\newblock {\em The Lancet},  {\it 396}, 10267, 2006--2017.
\newblock Publisher: Elsevier.

\bibitem[Dodaro et~al., 2021]{DBLP:journals/tplp/DodaroGGMMP21}
{\sc Dodaro, C.}, {\sc Galat{\`{a}}, G.}, {\sc Grioni, A.}, {\sc Maratea, M.},
  {\sc Mochi, M.}, {\sc and} {\sc Porro, I.} 2021.
\newblock An {ASP}-based solution to the chemotherapy treatment scheduling
  problem.
\newblock {\em Theory and Practice of Logic Programming},  {\it 21}, 6,
  835--851.

\bibitem[Dodaro and Maratea, 2017]{DBLP:conf/lpnmr/DodaroM17}
{\sc Dodaro, C.} {\sc and} {\sc Maratea, M.}
\newblock Nurse scheduling via answer set programming.
\newblock In {\sc Balduccini, M.} {\sc and} {\sc Janhunen, T.}, editors, {\em
  Proceedings of the 14th International Conference on Logic Programming and
  Nonmonotonic Reasoning ({LPNMR} 2017)} 2017, volume 10377 of {\em Lecture
  Notes in Computer Science}, pp. 301--307. Springer.

\bibitem[Erdem et~al., 2016]{DBLP:journals/aim/ErdemGL16}
{\sc Erdem, E.}, {\sc Gelfond, M.}, {\sc and} {\sc Leone, N.} 2016.
\newblock Applications of answer set programming.
\newblock {\em {AI} Magazine},  {\it 37}, 3, 53--68.

\bibitem[Gebser et~al., 2016]{DBLP:conf/iclp/GebserKKOSW16}
{\sc Gebser, M.}, {\sc Kaminski, R.}, {\sc Kaufmann, B.}, {\sc Ostrowski, M.},
  {\sc Schaub, T.}, {\sc and} {\sc Wanko, P.}
\newblock Theory solving made easy with clingo 5.
\newblock In {\sc Carro, M.}, {\sc King, A.}, {\sc Saeedloei, N.}, {\sc and}
  {\sc Vos, M.~D.}, editors, {\em Proceedings of {ICLP} (Technical
  Communications)} 2016, volume~52 of {\em {OASICS}}, pp. 2:1--2:15. Schloss
  Dagstuhl - Leibniz-Zentrum fuer Informatik.

\bibitem[Gebser et~al., 2015]{DBLP:conf/lpnmr/GebserKK0S15}
{\sc Gebser, M.}, {\sc Kaminski, R.}, {\sc Kaufmann, B.}, {\sc Romero, J.},
  {\sc and} {\sc Schaub, T.}
\newblock {Progress in clasp Series 3}.
\newblock In {\em {LPNMR}} 2015, volume 9345 of {\em {LNCS}}, pp. 368--383.
  Springer.

\bibitem[Gebser et~al., 2012]{DBLP:journals/ai/GebserKS12}
{\sc Gebser, M.}, {\sc Kaufmann, B.}, {\sc and} {\sc Schaub, T.} 2012.
\newblock Conflict-driven answer set solving: From theory to practice.
\newblock {\em Artificial Intelligence},  {\it 187}, 52--89.

\bibitem[Gebser et~al., 2018]{DBLP:journals/tplp/GebserOSR18}
{\sc Gebser, M.}, {\sc Obermeier, P.}, {\sc Schaub, T.}, {\sc
  Ratsch{-}Heitmann, M.}, {\sc and} {\sc Runge, M.} 2018.
\newblock Routing driverless transport vehicles in car assembly with answer set
  programming.
\newblock {\em Theory Practice of Logic Programming},  {\it 18}, 3-4, 520--534.

\bibitem[Gelfond and Lifschitz, 1991]{DBLP:journals/ngc/GelfondL91}
{\sc Gelfond, M.} {\sc and} {\sc Lifschitz, V.} 1991.
\newblock {Classical Negation in Logic Programs and Disjunctive Databases}.
\newblock {\em New Generation Computing},  {\it 9}, 3/4, 365--386.

\bibitem[{Gurobi Optimization, LLC}, 2021]{gurobi}
{\sc {Gurobi Optimization, LLC}} 2021.
\newblock {Gurobi Optimizer Reference Manual}.

\bibitem[Huang et~al., 2012]{Huang2012DecisionSS}
{\sc Huang, Y.-C.}, {\sc Zheng, J.-N.}, {\sc and} {\sc Chien, C.-F.} 2012.
\newblock Decision support system for rehabilitation scheduling to enhance the
  service quality and the effectiveness of hospital resource management.
\newblock {\em Journal of the Chinese Institute of Industrial Engineers},  {\it
  29}, 348 -- 363.

\bibitem[Huynh et~al., 2018]{huynh_hybrid_2018}
{\sc Huynh, N.-T.}, {\sc Huang, Y.-C.}, {\sc and} {\sc Chien, C.-F.} 2018.
\newblock A hybrid genetic algorithm with {2D} encoding for the scheduling of
  rehabilitation patients.
\newblock {\em Computers \& Industrial Engineering},  {\it 125}, 221--231.

\bibitem[Ignatiev et~al., 2019]{DBLP:journals/jsat/IgnatievMM19}
{\sc Ignatiev, A.}, {\sc Morgado, A.}, {\sc and} {\sc Marques{-}Silva, J.}
  2019.
\newblock {RC2:} an efficient maxsat solver.
\newblock {\em Journal of Satisfiability, Boolean Modeling, and Computation},
  {\it 11}, 1, 53--64.

\bibitem[Li and Chen, 2021]{Li_2021}
{\sc Li, X.} {\sc and} {\sc Chen, H.} 2021.
\newblock Physical therapy scheduling of inpatients based on improved genetic
  algorithm.
\newblock {\em Journal of Physics: Conference Series},  {\it 1848}, 1, 012009.

\bibitem[Martins et~al., 2014]{DBLP:conf/sat/MartinsML14}
{\sc Martins, R.}, {\sc Manquinho, V.~M.}, {\sc and} {\sc Lynce, I.}
\newblock Open-wbo: {A} modular maxsat solver,.
\newblock In {\em {SAT} 2014} 2014, volume 8561 of {\em LNCS}, pp. 438--445.
  Springer.

\bibitem[Morgado et~al., 2014]{DBLP:conf/cp/MorgadoDM14}
{\sc Morgado, A.}, {\sc Dodaro, C.}, {\sc and} {\sc Marques{-}Silva, J.}
\newblock {Core-Guided MaxSAT with Soft Cardinality Constraints}.
\newblock In {\em Proceedings of Principles and Practice of Constraint
  Programming - 20th International Conference, {CP} 2014} 2014, pp. 564--573,
  Lyon, France. Springer.

\bibitem[Niemel{\"{a}}, 1999]{DBLP:journals/amai/Niemela99}
{\sc Niemel{\"{a}}, I.} 1999.
\newblock {Logic Programs with Stable Model Semantics as a Constraint
  Programming Paradigm}.
\newblock {\em Annals of Mathematics and Artificial Intelligence},  {\it 25},
  3-4, 241--273.

\bibitem[{Olivier Roussel and Vasco Manquinho}, 2012]{pbcompetition}
{\sc {Olivier Roussel and Vasco Manquinho}} 2012.
\newblock {Input/Output Format and Solver Requirements for the Competitions of
  Pseudo-Boolean Solvers}.

\bibitem[Quinlan, 1986]{quinlan1986induction}
{\sc Quinlan, J.~R.} 1986.
\newblock Induction of decision trees.
\newblock {\em Machine learning},  {\it 1}, 1, 81--106.

\bibitem[Ricca et~al., 2012]{DBLP:journals/tplp/RiccaGAMLIL12}
{\sc Ricca, F.}, {\sc Grasso, G.}, {\sc Alviano, M.}, {\sc Manna, M.}, {\sc
  Lio, V.}, {\sc Iiritano, S.}, {\sc and} {\sc Leone, N.} 2012.
\newblock Team-building with answer set programming in the {G}ioia-{T}auro
  seaport.
\newblock {\em Theory and Practice of Logic Programming},  {\it 12}, 3,
  361--381.

\bibitem[Saikko et~al., 2016]{DBLP:conf/sat/SaikkoBJ16}
{\sc Saikko, P.}, {\sc Berg, J.}, {\sc and} {\sc J{\"{a}}rvisalo, M.}
\newblock {LMHS:} {A} {SAT-IP} hybrid maxsat solver.
\newblock In {\em {SAT} 2016} 2016, volume 9710 of {\em LNCS}, pp. 539--546.
  Springer.

\bibitem[Saverino et~al., 2021]{PMID:33995247}
{\sc Saverino, A.}, {\sc Baiardi, P.}, {\sc Galata, G.}, {\sc Pedemonte, G.},
  {\sc Vassallo, C.}, {\sc and} {\sc Pistarini, C.} 2021.
\newblock The challenge of reorganizing rehabilitation services at the time of
  covid-19 pandemic: A new digital and artificial intelligence platform to
  support team work in planning and delivering safe and high quality care.
\newblock {\em Frontiers in neurology},  {\it 12}, 643251.

\bibitem[Schimmelpfeng et~al., 2012]{schimmelpfeng_decision_2012}
{\sc Schimmelpfeng, K.}, {\sc Helber, S.}, {\sc and} {\sc Kasper, S.} 2012.
\newblock Decision support for rehabilitation hospital scheduling.
\newblock {\em OR Spectrum},  {\it 34}, 2, 461--489.

\end{thebibliography}

\end{document}